\documentclass[twocolumn]{aa}
\usepackage{epsfig}
\usepackage{lscape}
\usepackage{longtable}
\def\ar{{$\alpha_{\rm r}$~}}
\def\armu{{$\alpha_{\rm 1.4-41GHz}$~}}

\newcommand{\lsim}{{\lower.5ex\hbox{$\; \buildrel < \over \sim \;$}}}
\newcommand{\gsim}{{\lower.5ex\hbox{$\; \buildrel > \over \sim \;$}}}

\def\ee{\end{equation}}
\def\be{\begin{equation}}

\def\simlt{\ \raise -2.truept\hbox{\rlap{\hbox{$\sim$}}\raise5.truept   %
\hbox{$<$}\ }}
\def\simgt{\ \raise -2.truept\hbox{\rlap{\hbox{$\sim$}}\raise5.truept   %
\hbox{$>$}\ }}                                                          %

\def\be{\begin{equation}}
\def\ee{\end{equation}}
\def\newline{\hfil\break}

\def\la{\mathrel{\hbox{\rlap{\hbox{\lower4pt\hbox{$\sim$}}}\hbox{$<$}}}}
\def\ga{\mathrel{\hbox{\rlap{\hbox{\lower4pt\hbox{$\sim$}}}\hbox{$>$}}}}

\begin{document}

\title{The number counts, luminosity functions and evolution
of microwave-selected (WMAP) blazars and radio galaxies}
\author{Paolo~Giommi\inst{1}$^,$\inst{2},
        Sergio~Colafrancesco\inst{1}$^,$\inst{2}$^,$\inst{3},
        Paolo~Padovani\inst{4},
        Dario~Gasparrini \inst{1}$^,$\inst{5}$^,$\inst{6},
        Elisabetta~Cavazzuti\inst{1}$^,$\inst{2},
        Sara~Cutini\inst{1}$^,$\inst{5}$^,$\inst{6} }
 \institute{
            ASI Science Data Center, ASDC c/o ESRIN,
            Via G. Galilei 00044 Frascati, Italy.
\and
            Agenzia Spaziale Italiana,
            Unit\`a Osservazione dell'Universo,
            Viale Liegi 26 00198 Roma, Italy
\and
            on leave from: INAF - Osservatorio Astronomico di Roma,
            Via Frascati 33, 00040 Monteporzio, Italy
\and
            European Organisation for Astronomical Research in the
            Southern Hemisphere (ESO), Karl-Schwarzschild-Str. 2,
            D-85748 Garching bei M\"unchen, Germany
\and
            INAF personnell resident at ASDC under ASI contract I/024/05/1
\and
            Department of Physics, University of Perugia, Via A. Pascoli, I-06123, Perugia,
            Italy
 }

\offprints{Paolo Giommi paolo.giommi@asi.it}
\date{Received ; Accepted }

\authorrunning{Giommi et al.}
\titlerunning{ A WMAP complete sample of blazars}

\abstract{We have carried out an extensive search to identify the
counterparts of all the microwave foreground sources listed in the
WMAP 3-year catalogue using literature and archival data. Our work
has led to the identification of 309 WMAP sources, 98\% of which
are blazars, radio quasars or radio galaxies. Only 7 WMAP
detections were identified with other types of cosmic sources (3
starburst galaxies and 4 planetary/LBN nebulae). At present, 15
objects ($ < 5$\%) still remain without identification due to the
unavailability of optical spectroscopic data or a clear radio
counterpart. Our results allow us to define a flux limited sample
of 203 high Galactic latitude microwave sources ($f_{41GHz} \ge 1$
Jy, $|b_{\rm II}| > 15^\circ$) which is virtually completely
identified (99\%). The microwave band is ideally suited for blazar
statistical studies since this is the part of the electromagnetic
spectrum that is least affected by the superposition of spectral
components of different origin, and therefore by selection
effects. Using this data-set we have derived number counts,
luminosity functions and cosmological evolution of blazars and
radio galaxies at microwave frequencies. Our results are in good
agreement with those found at radio (cm) frequencies. The 5 GHz
bivariate blazar luminosity functions are similar to those derived
from the DXRBS survey, which shows that this sample is
representative of the blazar population at 41 GHz.
Microwave selected broad-lined quasars are about six times more
abundant than BL Lacs, a ratio that is similar to, or larger than,
that seen at radio and gamma-ray frequencies, once spectral
selection effects are taken into account. This strongly suggests
that  the mechanism responsible for the generation of gamma-rays
is, at first order, the same in all blazar types, leaving little
room for models (like external Compton radiation) that predict
very different gamma-ray emission in broad-lined and lineless
blazars. Our results confirm, and strengthen on a more solid
statistical base, the findings of Giommi \& Colafrancesco (2004,
2006) that blazars and radio galaxies are the largest contaminants
of the CMB anisotropy maps. We predict that these sources are also
bright gamma-ray sources, most of which will be detected by the
AGILE and {\it Fermi} satellites.}

\maketitle



\section{Introduction}

Blazars are the rarest and most peculiar type of Active Galactic Nuclei (AGN). So far
almost 3,000 such objects have been reported in the literature (see e.g., Massaro et al.
2009 for an updated catalogue) but their number is bound to increase significantly in the near future
as new, previously almost unexplored energy windows (e.g. microwave, hard X-ray, gamma-ray, TeV) are
becoming available for deep astronomical observations.

Blazar observational properties typically include irregular, sometimes large, rapid
variability, apparent super-luminal motion, flat radio spectrum, large and variable
polarization at radio and, especially, optical frequencies. Because of their special
observational properties, blazars are assumed to be sources emitting a continuum of
electromagnetic radiation from a relativistic jet that is viewed closely along
the line of sight thus causing strong relativistic amplification (e.g., \cite{bla78,Urry95}).

Blazars are a small fraction of all extragalactic sources but, unlike most other objects,
they are strong emitters across the entire electromagnetic spectrum. In the optical and
soft X-ray bands -- where the radiation that we observe is mostly due to thermal emission
originated in stars and galaxies or from the accretion process onto the central engine of
AGN -- blazars are a tiny minority, but in other parts of the electromagnetic spectrum,
where thermal emission becomes unimportant, they are often the dominant population in the
extragalactic sky. Following the technological evolution and the availability of
astronomical resources, most blazars have so far been discovered as counterparts of flat
spectrum radio emitters or as X-ray sources. Flux limited samples have been compiled from
several surveys in both bands (e.g., \cite{Pad07a} and references therein).
In both these spectral regions, however, there are complications that may lead
to selection effects of different nature: contamination by extended radio
emission from the radio-lobes at cm wavelengths and contamination by radiation
produced in the accretion process in the soft X-rays, among others. This may be
particularly important in medium-low luminosity Flat Spectrum Radio Quasars
(FSRQs) where the thermal emitted power from accretion processes and the
non-thermal emission processes may be of comparable importance (e.g.
\cite{Landt08}). The level of both thermal and non-thermal components can be in
fact comparable also in the case of bright objects like, e.g., 3C 273 (see
\cite{Grandi04}).

The spectral energy distribution (SED) of blazars includes a synchrotron
low-frequency component that peaks (in a $Log(\nu f(\nu))-Log(\nu)$
representation) between the far infrared and the X-ray band, followed by an
Inverse Compton high-frequency component that has its maximum in the hard X-ray
band or at higher energies (see, e.g. \cite{roxa0810}), depending on the
location of the synchrotron peak, and extends into the $\gamma$-ray or even in
the TeV band.

Only recently the microwave region of the electromagnetic spectrum has become
available to allow for systematic studies of blazars over large cosmological
volumes. This frequency band is indeed particularly suited for the selection of
blazars since at these frequencies the contamination from radio extended
components with steep spectra is no longer present and the emission from the
accretion process is negligible.

In this paper we present the first flux-limited sample of microwave selected blazars
extracted from the catalogue of bright foreground sources detected by the WMAP satellite
(\cite{bennett03,Hinshaw07}). Using this statistically complete sample we derive the
number counts, the luminosity functions and the cosmological evolution of blazars and
radio galaxies in the microwave band and we compare these properties with those observed
at cm frequencies.

Throughout this paper, source spectra are written as $S_{\nu} \propto \nu^{-\alpha}$
where $\alpha$ is the spectral index. We use a flat, vacuum-dominated CDM cosmological
model with $H_0 = 70$ km s$^{-1}$ Mpc$^{-1}$, $\Omega_{\rm M} = 0.3$, and $\Omega_{\rm
\Lambda} = 0.7$ (\cite{cosmopar03}). To compare some of our results with previous works,
we will also adopt a value $H_0 = 50$ km s$^{-1}$ Mpc$^{-1}$ and an empty universe
cosmology with $\Omega_{\rm M} = 0$, and $\Omega_{\rm \Lambda} = 0$, as specifically
declared in the text.

\section{The sample}

The catalogue of WMAP bright foreground sources that we consider
here is based on the WMAP 3-year data (\cite{Hinshaw07}) and
includes 324 sources detected by WMAP in at least one channel
after three years of operation. Because of the higher sensitivity
of the 3-year maps this catalogue is a significant improvement
compared to that based on the first year data (\cite{bennett03})
that included 208 sources.

For the purpose of this paper we will define our flux limited sample using the source
fluxes as observed in the WMAP 41 GHz channel as a compromise between sensitivity and
completeness, and the need to use the highest frequency band for the definition of a
complete sample suitable for statistical purposes.

To arrive at the definition of such a complete sample, we first discuss the effect of
source confusion in the estimate of the 41 GHz flux of the objects likely associated to
the WMAP foreground sources. Then, we provide a classification scheme for each object
associated to the WMAP source and finally we derive the statistical properties of our
complete sample.

\subsection{Source confusion}

The WMAP experiment has a limited angular resolution which ranges
between $\sim 0.93$ deg to $0.23$ deg (FWHM) for the frequency
bands 23 GHz to 94 GHz\footnote{
http://map.gsfc.nasa.gov/m$_-$mm/ob$_-$techres.html}. In
particular, the 41 GHz WMAP channel, to which we refer in our
study, has an angular resolution of $31.8$ arcmin (FWHM). Although
this limited resolution, combined with the surface density of
radio sources at around 1 Jy, does not imply a large fraction of
confused sources,
in some cases the probability that more than one object
(especially those with flat radio spectra, like blazars) falls
within the WMAP 41 GHz beam may not be negligible. When this
happens the microwave flux attributed to a WMAP foreground source
may be contaminated.
We analyzed all the WMAP beams centered on the point-like sources
detected at 41 GHz in order to find the level of confusion
produced by the presence of multiple radio sources in the same
beam, taking into consideration the effect of primary beam
attenuation (Page et al. 2003). To this purpose, we adopted the
following procedure:
\begin{itemize}
\item we started our analysis by associating the WMAP foreground source to the radio source
listed in the WMAP 3 year catalogue (\cite{Hinshaw07}). However, in some cases
we found arguments to propose a different counterpart, as specifically
discussed in Sect. 2.3 below;
\item we choose to select radio sources potentially contaminating the 41 GHz flux as
all sources within the WMAP 41 GHz beam centered on the object
associated to the WMAP source that have radio flux  at 5 GHz,
$f_{5GHz} > 100$ mJy;
\item for each radio source detected at 5 GHz within the WMAP beam, we estimated its 41 GHz
flux by extrapolating the observed flux at lower frequencies using
a linear regression (i.e., a power-law spectrum) technique. To
this aim we used a variety of (non-simultaneous) radio catalogues
including SUMSS, NVSS, GB6, PMN, and CRATES.
Finally, we corrected the central WMAP source flux by subtracting
the 41 GHz flux contributions of the possible contaminating
objects taking into account the distance-dependent primary beam
attenuation of the WMAP 41 GHz channel.

As the minimum error on the WMAP flux is 0.1 Jy, the correction
was applied only when larger that this value. Our method could
suffer from variability of the contaminants, which are mostly
flat-spectrum radio sources, but this is unlikely to be a major
problem because of the small fraction of ''corrected'' sources and
amount of contamination. The percentage of sources, which ended up
having their 41 GHz flux corrected, in fact, is only $\sim 7\%$,
with the mean correction being $\sim 13\%$.
\end{itemize}

\begin{figure}[!h]
\begin{center}
\includegraphics[height=6.cm,angle=0]{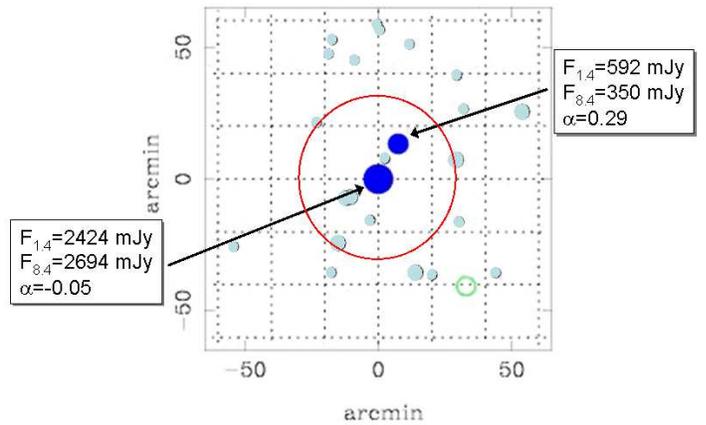}
\end{center}
\caption{An example of source confusion in the field of WMAP3
J0339-0144. Light grey filled circles represent radio sources
(NVSS) either with steep radio spectrum or without spectral
information; flat-spectrum radio sources are shown as dark blue
filled circles. The green open circle marks the galaxy cluster
ZW0334-0237 which is located 52 arcmin away from the center of the
field. The size of the gray and blue circles is proportional to
1.4 GHz radio flux of the source they represent.}
 \label{fig.wmap3j0339}
\end{figure}

An example of a WMAP source which results confused by another flat-spectrum
radio source in the field is shown in Fig. \ref{fig.wmap3j0339}.
This field includes two flat spectrum radio sources in the WMAP
41-GHz channel beam (the red circle with radius FWHM = 31.8
arcmin): the central source is located 1.8 arcsecs from the center
of the field and has a flat spectrum with slope $\alpha_{1.4-4.8
GHz}= -0.05$; there is a second bright flat spectrum source at 15
arcmin from the center of the field.

\begin{figure}[!h]
\begin{center}
\includegraphics[height=7.cm,angle=0]{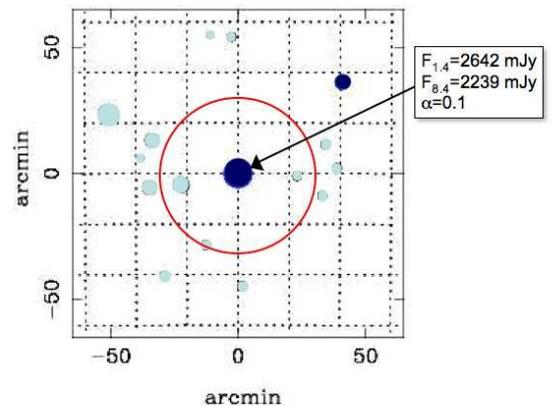}
\end{center}
\caption{Same as Fig. \ref{fig.wmap3j0339} but for the field of  WMAP3
J2348-1631. The various sources in the field are NVSS sources either with steep
radio spectrum or without spectral information (gray open circles) and
flat-spectrum radio sources (blue open circles). The size of the gray and blue
circles is proportional to 1.4 GHz radio flux. The WMAP source is clearly
identified with the bright flat spectrum radio source (the blazar PKS~2345-16)
at the center of the field.}
 \label{fig.wmap3ok}
\end{figure}

Fig. \ref{fig.wmap3ok} shows a typical case in which there is no doubt about the
association between the WMAP source and its radio counterpart.
In this case, there is only one flat-spectrum ($\alpha_{1.4-4.8 GHz}= 0.1$)
radio source in the WMAP 41-GHz channel beam and it is very close (1.8 arcsecs)
to the center of the field.

\subsection{Object classification}

We follow standard criteria and classify our sources based on
  their optical and radio spectra. The blazar class includes BL Lacertae
  objects, historically characterized by an almost complete lack of
  emission lines, and FSRQs, which by definition display broad, strong
  emission lines. A dividing value of full-width half maximum $FWHM = 1000$
  km/s between ``narrow'' and ``broad" emission lines is typically used.
  BL Lacs and radio galaxies can both have narrow lines so their separation
  is typically done on the basis of the value of the Ca break, a stellar
  absorption feature in the optical spectrum defined by $C = (f_+ - f_-) /
  f_+$ (where $f_-$ and $f_+$ are the fluxes in the rest frame wavelength
  regions $3750 - 3950$ \AA~ and $4050 - 4250$~\AA~ respectively). A
  separation value of $C=0.4$ is normally adopted (e.g., \cite{Pad07a}), as
  sources with $C < 0.4$ are believed to be dominated by non-stellar emission.  We
  also make the commonly accepted distinction between steep spectrum radio
  quasars (SSRQ), $\alpha_{\rm r} > 0.5$, and FSRQ, $\alpha_{\rm r} \le
  0.5$. In order to separate BL Lacs (with broad emission lines) from
  radio quasars a limit of 5 \AA~on the rest frame equivalent width of any
  detected emission line is normally used (e.g., \cite{Sti91}).

  As most of the objects in the WMAP 3-year catalogue are bright and well
  known radio sources, the literature is rich with data. We have used the
  broad-band SED as well as any optical spectroscopy information available
  in the literature (mostly taken from the recent blazar catalog of
  \cite{Massaroetal2009}, from the NASA Extragalactic Database and from the SDSS-DR6 on-line services)
  to classify all sources in the WMAP 3-year catalogue.

  In summary, a WMAP source is associated to:
\begin{itemize}
\item an FSRQ if its spectral index at cm frequencies \ar and between the
radio and the microwave region \armu is flatter than 0.5, and if its optical spectrum
shows strong and broad emission lines typical of QSOs, as detailed above;
\item a BL Lac if its spectral index at cm frequencies \ar and between the
radio and the microwave region \armu is flatter than 0.5 and if its optical spectrum does
not show strong broad emission lines and looks non-thermal, as detailed above;
\item an SSRQ if its spectral index at cm frequencies \ar
is steeper than 0.5 (independently of the \armu index, which could
be significantly flatter than \ar) and if its optical spectrum
shows strong and broad emission lines;
\item a radio galaxy if its spectral index at cm frequencies \ar
is steeper than 0.5 (independently of the \armu index), its radio emission is clearly
extended, it has weak, narrow lines and an optical spectrum dominated by stellar
emission, and if the host galaxy is clearly visible at optical/IR frequencies.
\end{itemize}

In a few cases we could not find sufficient information to classify a
source with blazar characteristics as an object belonging to one of the
above categories. In the following these sources are labelled as Blazars -
Unknown type. Finally, in three cases the WMAP source appears to be
classified in the literature as a starburst galaxy.

High frequency surveys detect a population of GigaHertz Peaked
Spectrum (GPS) sources (e.g., \cite{odea98}). We have
cross-correlated the WMAP list with the recent compilation by
\cite{Labiano07} and found 11 GPS sources, 5 of which belonging to
the complete sample. These objects are marked accordingly in Tab.
\ref{tab:list}.

\subsection{Notes on individual objects}

\newline {\bf WMAP3~J0026-3511} {\bf (complete sample).}
There are no optical spectra of this object available in the
literature. Its very flat radio spectrum and its blazar-like SED
make this source a high-confidence blazar candidate. This source
is very weak at low frequencies with a flux of only 14 mJy at 843
MHz (from SUMSS) and 25 mJy at 1.4 GHz  (from NVSS). It is listed
in the CRATES catalogue (\cite{crates}) as a very inverted radio
source (spectral slope \ar = - 1.4) with a flux of 121 and 314 mJy
at 4.8 and 8.4 GHz, respectively.

\newline {\bf WMAP3~J0210-5100} {\bf (complete sample).} This source is a transition object between FSRQs and BL Lac (\cite{Massaroetal2009}). Given the relatively high redshift of this object
($z=1$), we include it in the FSRQ sample.

\newline {\bf WMAP3~J0540-5416, J0550-5732, J0633-2217, J1038+0511,
J1102-4400, J1333+2723, J2035-6845 and J2148-7757} {\bf (complete sample: 1st, 2nd, and 4th source).}
These are all FSRQs that have been very recently identified
by \cite{cratesb}  as part of the identification program of the CGRaBS
gamma-ray source candidates.

\newline {\bf WMAP3~J1149-7932}. The object associated to this source in the WMAP 3 year catalog
is not the most likely counterpart since it is too faint and with a rather steep
spectrum.

\newline {\bf WMAP3~J1227+1124}. The position of the WMAP source does not correspond to
a definite radio source. The GB6 source in the field is rather faint (34 mJy) and with
flat spectrum. There is, however, a bright radio source associated to an interactive
radio-galaxy (VPCX 27) at $\approx 25$ arcmin from the nominal position of the WMAP 3
year source.

\newline {\bf WMAP3~J1231+1351}. Hinshaw et al. (2007) associate this source to
GB6 J1231+1344 which is located 10 arcmin away from the WMAP source centroid. However,
this radio source has a steep radio spectrum, while the source GB6 J1232+1359, at about
20 arcminutes from the WMAP centroid, is flat and brighter then GB6 J1231+1344. For this
reason we associate WMAP3~J1231+1351 to GB6~J1232+1359.

\newline {\bf WMAP3~J1305-4928}.
The WMAP source is associated to the galaxy NGC~4945, one of the four brightest FIR
sources in the sky outside the local group. NGC~4945 hosts both vigorous nuclear star
formation (Moorwood \& Oliva 1994, Spoon et al. 2000) and a peculiar broad-lined AGN with an
estimated bolometric luminosity of $\sim ~60\%$ of the nuclear FIR luminosity. We
classified this object as FIR-Starburst galaxy in our sample.

\newline {\bf WMAP3~J1657+4749}.  We found 2 objects located inside the WMAP 41 GHz beam: 4C 48.41 and [HB89]
1656+477, both with a flat radio spectrum. The WMAP source has a
41 GHz flux of 0.6 Jy, but the two sources mentioned above show a
predicted flux at 41 GHz of 0.5 Jy and 1.3 Jy, respectively.
Hinshaw et al. (2007) associate this source to the object GB6 1658+4737, that is [HB89]
1656+477. We agree with this preliminary association.

\newline {\bf WMAP3~J1924-2914} {\bf (complete sample).} Despite this source (PKS J1924-2914, PMN J1924-2914 or OV-236)
being one of the brightest extragalactic objects at mm frequencies, it was not
included in the WMAP 3-year catalog of Hinshaw et al. (2007) because it falls
within the Kp0 cut that is used to avoid including a large number of Galactic
sources in the catalog. Since the Kp0 cut is based on WMAP K-band intensity and
this source is so bright, it falls within that cut even though it is a
relatively high latitude source ($b_{\rm II} = -19.6^\circ$;
this is the only high-latitude extragalactic
source to suffer this fate: G. Hinshaw private communication (see also Fig. 2
of Bennett et al. 2003).
PKS J1924-2914, a well known, bright FSRQ, was clearly detected in
the WMAP maps in all bands. Therefore, we have included it in our
blazar sample. We estimated (from an analysis of the WMAP 3-yr
data) that this source has a flux of $f_{41GHz}= 13.29 \pm 1.12$
Jy (with a mm spectral index $\alpha = 0.6$) and so it has been
included in our complete sample.

\newline {\bf WMAP3~J2333-2340} {\bf (complete sample).} The optical spectrum of this source (Wilkes et al. 1983) shows
emission lines and therefore, for consistency with current literature classification
methods, we classify it as a radio galaxy. However the SED is typical of a blazar at all
frequencies and shows a large optical variability. In a high optical state this object
would have been classified as a BL Lac.

\subsection{The complete sample}

The full list of WMAP sources is given in Table \ref{tab:list}
where column 1 gives the WMAP source name following the IAU naming
convention, column 2 gives the WMAP number defined in the first
year catalogue (if available), column 3 gives a common name (e.g.
NGC, PKS, 3C etc.) if the source is a previously known object;
columns 4 and 5 give the Right Ascension and Declination (J2000.0)
of the counterpart of the WMAP source; column 6 gives the source
class following the classification scheme described in Sect. 2.2;
column 7 gives the redshift when available; columns 8 and 9 give
the flux  at 5 GHz (from the literature) and at 41 GHz (from the
WMAP-3 yr catalogue); column 10 gives the 41 GHz flux corrected
for source confusion as described in Sect.2.1 and column 11 states
if the source is part of the flux limited, complete sample.

\begin{figure}[!h]
\includegraphics[height=8.cm,angle=-90]{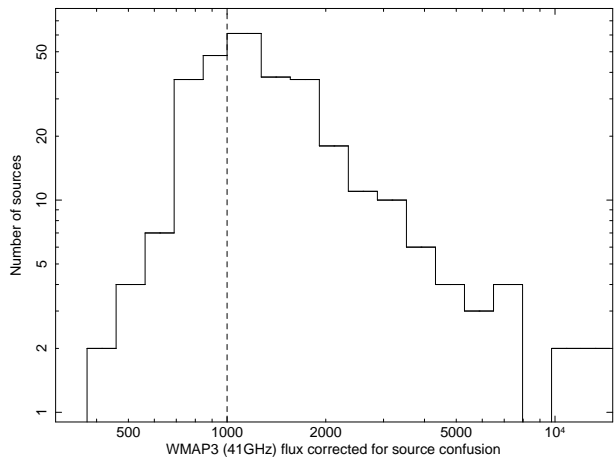}
\caption{The distribution of the corrected 41 GHz flux  in the
entire WMAP sample. The sample appears to be clearly incomplete
below 1 Jy, as marked by the dashed vertical line.}
 \label{fig.flux_histo}
\end{figure}

The flux list of WMAP sources and the limited sample that we use
for our statistical analysis is defined as the sample of all
sources in the WMAP 3-yr catalogue with flux  at 41 GHz larger
than 1 Jy and with Galactic latitude $|b_{\rm II}| > 15^\circ$.
The latter condition is necessary to limit source confusion and
the complications associated to the emission close the Galactic
plane. Taking also into account the higher latitude regions
excluded by the Kp0 mask of \cite{bennett03} our total area is
28,457 square degrees.
We chose $f_{41GHz}=1$ Jy as the flux limit of our sample because
the number of sources below this flux value drops sharply as shown
in Fig. \ref{fig.flux_histo} where the distribution of 41 GHz
fluxes is plotted for the entire sample. Above $F_{41GHz}=1$ Jy
(marked by the vertical dashed line) the slope of the distribution
is steep and uniform indicating a very good level of completeness
even close to the flux  limit.

Table \ref{tab:stat} gives the statistics of WMAP sources identification both for the full and the
complete sample.
\begin{table}[!h]
\begin{center}
\begin{tabular}{lc}
\hline
   \multicolumn{1}{l}{Source type}
 & \multicolumn{1}{c}{Number in sample} \\
 & \multicolumn{1}{c}{total/complete} \\
\hline
 FSRQ & 202/137 \\
 BL Lac & 38/24 \\
 Blazars-unknown type & 17/11\\
 SSRQ& 14/9\\
 Radio galaxies& 31/15 \\
 Starburst galaxies & 3/2 \\
 Planetary/LBN nebulae & 4/2 \\
 Unidentified/no radio-counterpart & 15/3 \\
 Total & 324/203  \\
\hline
\end{tabular}
\caption{Summary of WMAP objects identification}
 \label{tab:stat}
\end{center}
\end{table}

\section{The blazar number counts at microwave frequencies (41 GHz)}

\begin{figure}[!h]
\begin{center}
\includegraphics[height=9.cm,angle=-90]{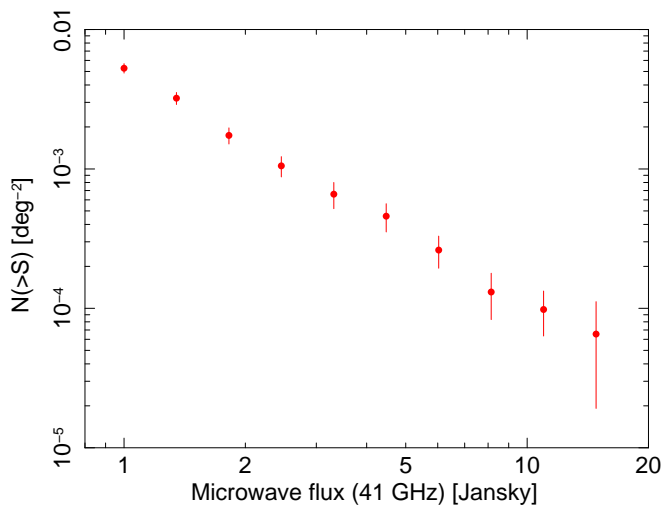}
\end{center}
\caption{The 41 GHz LogN-LogS of all blazars in the complete
sample ($|b_{\rm II}| > 15^\circ$) is shown by the filled dots
with error bars. For clarity only a subset of the points is shown.
}
 \label{fig.logns}
\end{figure}

In this section we present the 41 GHz logN-logS of blazars derived from our complete
sample and we compare it to the counts derived at radio (5 GHz) frequency.
Fig. \ref{fig.logns} shows the integral LogN-LogS of {\it all} blazars (i.e.
FSRQs, BL Lacs, and unclassified type) in the complete sample. The
blazar counts at 41 GHz are steep and well described by a simple power law of
the type $N(>S) \propto S^{-\beta}$. We use the maximum likelihood method
(\cite{cra70}) to estimate the slope of the number counts using all the
available information\footnote{This method operates on the
differential distribution and obviates to the fact that individual points in
integral counts are not independent.}. This method yields $\beta = 1.64 \pm
0.14$, somewhat steeper than the Euclidean value of 1.5. The slopes for blazar
sub-classes and other classes are consistent with this value within the
typically large error bars.

This 41 GHz logN-logS is in very good agreement both in slope and
normalization with the 5GHz blazar counts at flux densities $> 1$ Jy (see
e.g. Giommi et al. 2006). This is expected given the flat $\langle
\alpha_{\rm 5-41 GHz} \rangle = 0.02$ average spectral slope of our
blazars.

\section{Cosmological evolution}

The simplest way to study the evolutionary properties of a sample is through the
$V/V_{\rm max}$ test (\cite{Sch68}). Values of $\langle V/V_{\rm max} \rangle$
significantly different from 0.5 indicate evolution, which is positive (i.e., sources
were more luminous and/or more numerous in the past) for values $> 0.5$, or negative
(i.e., sources were less luminous and/or less numerous in the past) for values $< 0.5$.
One can also fit an evolutionary model to the sample by finding the evolutionary
parameter which provides $\langle V/V_{\rm max} \rangle = 0.5$.

We have computed the quantity $V/V_{\rm max}$ for our sources,
with statistical errors given by $\sigma = 1/\sqrt{12~N}$
(\cite{Avni80}). To have a simple estimate of the sample evolution
we have also derived the best fit parameter $\tau$ (given in units
of the age of the universe) of a pure luminosity evolution model,
i.e., $P(z) = P(0)\; {\rm exp}[T(z)/\tau]$, where $T(z)$ is the
a-dimensional look-back time: the smaller $\tau$ is the stronger
the evolution. We assume here that some luminosity evolution takes
place, based on previous studies in the radio and other bands
(e.g., \cite{Pad07a,Croom04}).

\begin{figure}[!h]
\begin{center}
\includegraphics[height=8.cm,angle=0]{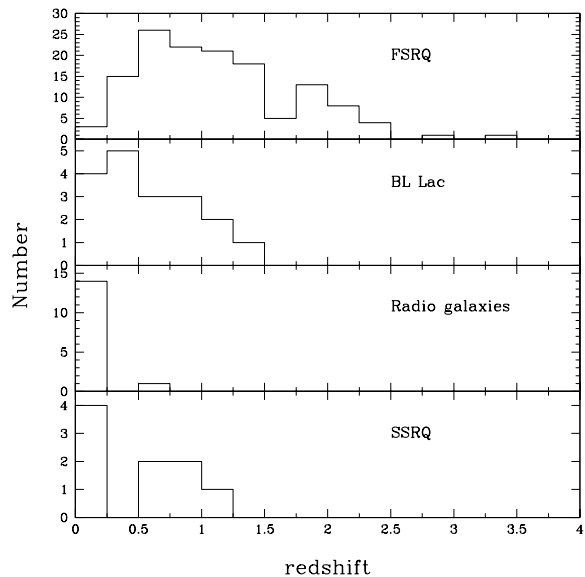}
\end{center}
\caption{The redshift distributions of blazars (FSRQs and BL Lacs), SSRQs, and
radio galaxies.}
\label{fig.zdist}
\end{figure}

The redshift distribution of WMAP blazars, SSRQs, radio galaxies is shown in
Fig. \ref{fig.zdist}.

The results of the cosmological evolution for our sample are shown in Table
\ref{tab:evol}, which lists the source sub-sample in column (1), the number of
sources in column (2), the mean redshift in column (3), the value of $\langle
V/V_{\rm max} \rangle$ and $\tau$ in columns (4) and (5) for our $\Lambda$CDM
reference cosmological model, and the value of $\langle V/V_{\rm max} \rangle$
and $\tau$ in columns (6) and (7) for an empty Universe cosmology, for
comparison with previous results.

The main results of our analysis can be summarized a follows:
\begin{enumerate}
\item FSRQs evolve at the $4.6\sigma$ level
($\langle V/V_{\rm max}\rangle= 0.614 \pm 0.025$),
confirming a well known result (e.g., \cite{P92,Urry95,Pad07a}). The
evolutionary parameter of this class is consistent with that derived by \cite{Urry95} for
the 2 Jy FSRQ sample in an empty Universe ($\tau = 0.23^{+0.07}_{-0.04}$), for the simple
case of pure luminosity evolution.
\item BL~Lacs display moderate evolution,
i.e., their value of $\langle V/V_{\rm max} \rangle = 0.62 \pm
0.06$ is different from 0.5 at the $2 \sigma$ level. The fact that
$25$\% of the sources have no redshift information is not much of
a problem, as redshift affects $V/V_{\rm max}$ values way less
than flux.
\item Non blazars are consistent with no evolution
(radio galaxies $\langle V/V_{\rm max} \rangle = 0.61 \pm 0.07$
and SSRQ $\langle V/V_{\rm max} \rangle = 0.63 \pm 0.10$).
However, this is likely due to small number statistics, as SSRQ
and high-power radio galaxies are known to evolve strongly in the
radio band (e.g., \cite{Urry95}).
\item Blazars of unknown type show hints of evolution ($\langle V/V_{\rm max}
\rangle = 0.70 \pm 0.09$) at the $2.3\sigma$ ($2.5\sigma$ for an empty Universe
cosmology) level, with a best fit value of $\tau$ consistent, within $\sim 1.5\sigma$, with
that of FSRQs.
\end{enumerate}

\begin{figure}[!h]
\begin{center}
\includegraphics[height=8.cm]{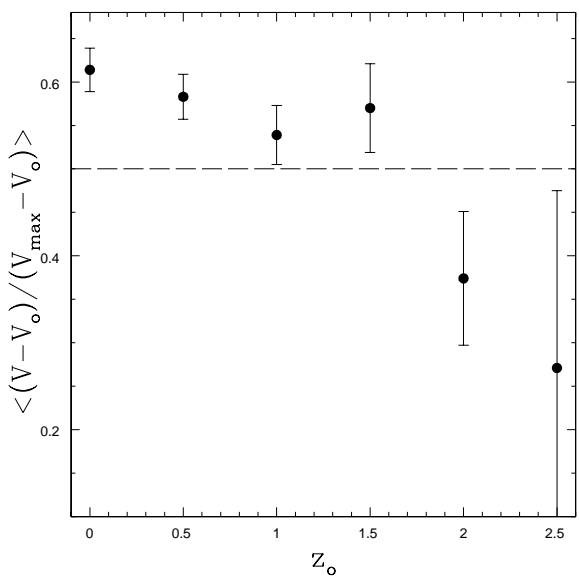}
\end{center}
\caption{The banded statistic, $\langle (V - V_o)/(V_{\rm max} - V_o) \rangle$ versus
$z_o$ for WMAP FSRQ. The horizontal dashed line indicates the value of 0.5 expected under
the null hypothesis of no evolution. } \label{fig.vovm_z}
\end{figure}

Given the available FSRQ statistics, we can move beyond the assumption of pure luminosity
evolution and study possible redshift dependencies. We do this by using the so-called
banded $\langle V/V_{\rm max} \rangle$ statistic, i.e. the quantity $\langle (V - V_{\rm
o})/(V_{\rm max} - V_{\rm o}) \rangle$, where $V_o$ is the cosmological volume enclosed
by a given redshift $z_o$ (\cite{Dunlop90}). This procedure allows the detection of any
high-redshift (possibly negative) evolution by separating it by the well-known strong
(positive) low-redshift evolution. Fig. \ref{fig.vovm_z} shows that $\langle (V - V_{\rm
o})/(V_{\rm max}  - V_{\rm o}) \rangle$ is basically constant up to $z \sim 1.5$ with a
significant drop at higher redshifts. Its value, in fact, goes from being $> 0.5$ at the
$4.6\sigma$ level at $z \sim 0$ to values $\le 0.5$ for $z \ga 2$. However, given that
only $\sim 10\%$ of the FSRQ are above this redshift, we assume a pure luminosity
evolution for the whole sample. This is confirmed, within the somewhat limited
statistics, by a study of the LF in redshift bins.
\begin{table*}[!h]
\caption{WMAP sample Evolutionary Properties}
\begin{tabular}{lrllclc}
\hline ~&~&~&\multispan2{$H_0 = 70$, $\Omega_{\rm M} = 0.3$, $\Omega_{\rm
 \Lambda} = 0.7$~~~~}& \multispan2{$H_0 = 50$, $\Omega_{\rm M} = 0$,
 $\Omega_{\rm \Lambda} = 0$}\\
  Class & Number & ~~~$\langle z \rangle$ & $\langle V/V_{\rm max} \rangle$
 & $\tau$  & $\langle V/V_{\rm max} \rangle$ & $\tau$ \\
\hline
 FSRQ & 137 & $1.13\pm0.05$ & $0.614\pm0.025$ & $0.35^{+0.07}_{-0.05}$
&$0.621\pm0.025$ & $0.26^{+0.05}_{-0.03}$\\
 BL Lacs & 24 & $0.56\pm0.09^a$ & $0.619\pm0.059^b$ & $0.31^{+0.21}_{-0.08}$
&$0.626\pm0.059^b$ & $0.24^{+0.15}_{-0.05}$\\
 Radio Galaxies & 15 & $0.08\pm0.04$ & $0.610\pm0.074$ & ...$^c$
&$0.610\pm0.074$ & ...$^c$\\
 Unclassified blazars & 11 & $0.77\pm0.16^d$ & $0.703\pm0.087^e$ & $0.22^{+0.11}_{-0.04}$
&$0.713\pm0.087^e$ & $0.17^{+0.07}_{-0.03}$ \\
 SSRQ & 9 & $0.50\pm0.14$ & $0.626\pm0.096$ & ...$^c$
&$0.632\pm0.096$ & ...$^c$\\
 \hline
 \multicolumn{5}{l}{\footnotesize $^a$Excluding the 6 sources without redshift}\\
 \multicolumn{5}{l}{\footnotesize $^b$Assuming $z = \langle z \rangle$ for the sources
 without redshift}\\
 \multicolumn{5}{l}{\footnotesize $^c$$\langle V/V_{\rm max} \rangle$ not significantly
 different ($< 2~\sigma$) from 0.5: no evolution assumed}\\
 \multicolumn{5}{l}{\footnotesize $^d$Excluding the 6 sources without redshift}\\
 \multicolumn{5}{l}{\footnotesize $^e$Assuming $z$  equal to the mean blazar redshift for
 the sources without redshift}\\
\end{tabular}
\label{tab:evol}
\end{table*}

\section{The luminosity function of WMAP blazars}

\subsection{BL Lacs}

\begin{figure}[!h]
\begin{center}
\includegraphics[height=8.cm]{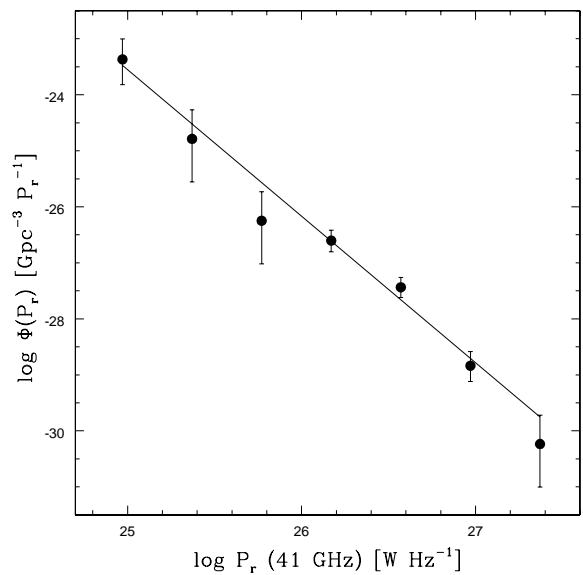}
\end{center}
\caption{The local, de-evolved 41 GHz luminosity function of BL Lacs (filled points with error bars).
Error bars correspond to $1\sigma$ Poisson errors (\cite{Gehrels86}). The solid line is a
weighted least-squares fit to the data of the form $\phi(P_{\rm r}) \propto P_{\rm
r}^{-2.62}$ (see text for details).}
 \label{fig.lf_bllacs}
\end{figure}

The local LF of WMAP BL Lacs is shown in Fig. \ref{fig.lf_bllacs}. Based on the $\langle
V/V_{\rm max} \rangle$ analysis, this has been de-evolved to zero
redshift using $\tau = 0.31$. We have assumed $z = \langle z \rangle \sim 0.56$
for the 6/24 BL Lacs without redshift.

We fitted the LF with a single power law $\phi(P_{\rm r}) \propto
P_{\rm r}^{-B_r}$. Varying the luminosity binning, the
differential slope is in the range $2.4 < B_r < 2.7$. For a
representative bin size of $\Delta \log P_r = 0.4$, a weighted
least-squares fit yields $\phi(P_{\rm r}) \propto P_{\rm
r}^{-2.62\pm0.19}$ ($\chi^2_{\nu} \sim 0.94$ for 5 degrees of
freedom). The total number density of BL Lacs in the luminosity
range $7 \times 10^{24} - 10^{28}$ W/Hz, derived independently of
bin size from the integral LF, equal to $\sum 1/V_{\rm max}$, is
$44\pm29$ Gpc$^{-3}$ (see eqs. 9 and 10 of \cite{con02}).

We note our assumption about the missing redshifts could somewhat
bias our LF. Therefore, we have checked this by assuming $z= 1$
for these sources. As discussed in \cite{Pad07a}, their
featureless continuum might suggest a relatively high redshift.
However, this assumption results in a LF, which is consistent with
the previous one well within the errors, with $\phi(P_{\rm r})
\propto P_{\rm r}^{-2.51\pm0.19}$ ($\chi^2_{\nu} \sim 0.69$ for 5
degrees of freedom).

\begin{figure}[!h]
\begin{center}
\includegraphics[height=8.cm]{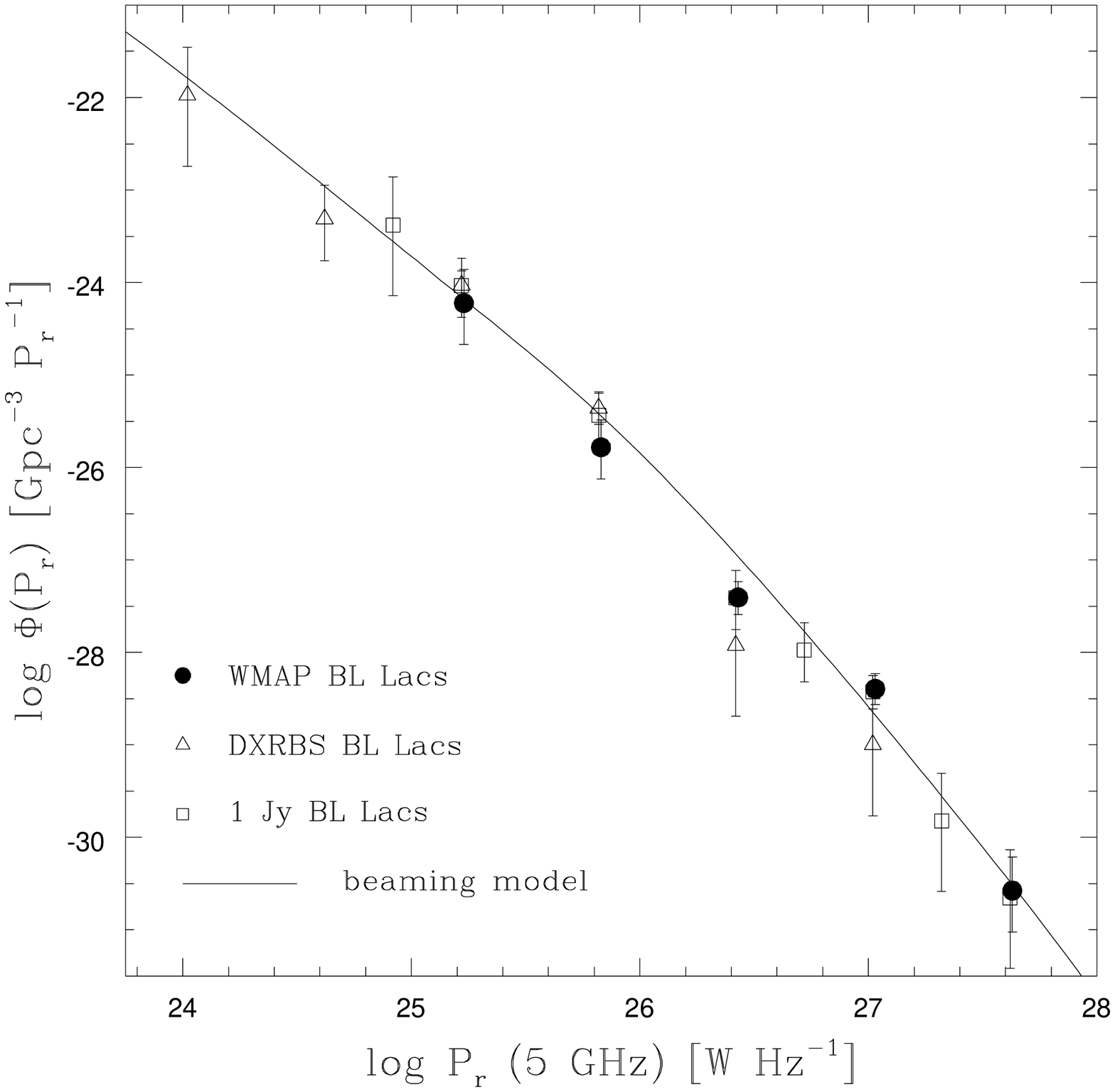}
\end{center}
\caption{The bivariate 5 GHz luminosity function of WMAP BL Lacs (filled points) and DXRBS (open
triangles) compared to the predictions of a beaming model based on the 1 Jy luminosity
function and evolution (solid line \cite{Urry95}). The open squares represent the 1 Jy
luminosity function \cite{Sti91}. Error bars correspond to $1\sigma$ Poisson errors
\cite{Gehrels86}. The DXRBS points have been corrected to take into account their
somewhat different definition of a BL Lac (\cite{Pad07a}). For consistency with
\cite{Sti91} and \cite{Urry95} the WMAP luminosity function has been de-evolved to zero
redshift and an $H_0 = 50$ km s$^{-1}$ Mpc$^{-1}$, $\Omega_{\rm M} = 0$, and $\Omega_{\rm
\Lambda} = 0$ cosmology has been adopted.} \label{fig.lf_bllacs_dxrbs}
\end{figure}

We have also derived the bivariate LF of WMAP BL Lacs at 5 GHz for
an empty Universe cosmology, to compare it with previous
determinations and with the predictions of unified schemes. This
is presented in Fig. \ref{fig.lf_bllacs_dxrbs} (filled points),
which shows also the DXRBS (open triangles, \cite{Pad07a}), 1 Jy
LF (open squares, \cite{Sti91}) and the predictions of a beaming
model based on the 1 Jy LFs and evolution (solid line,
\cite{Urry95}). These predictions assume that Fanaroff-Riley type
I (FR I; \cite{FR74}) galaxies are the counterparts of BL Lacs
with their jets on the plane of the sky and show what one should
expect to find when reaching powers lower than those used to
constrain the LF at the high end. For consistency with
\cite{Sti91} we have de-evolved the WMAP BL Lacs assuming $\tau =
0.32$, which is within $0.5\sigma$ from our own value (see Tab.
\ref{tab:evol}). We note that 14/34 of the 1 Jy BL Lacs belong
also to the WMAP sample.

The LF of WMAP BL Lacs is in very good agreement, in the region of
overlap, with the DXRBS and 1 Jy LFs. In particular, the WMAP and 1 Jy LFs cover roughly
the same power range, a result that is expected given the relatively similar flux limits.
The WMAP LF is also in good agreement with the predictions of unified schemes. For this
cosmology the total number density of BL Lacs in the luminosity range $2 \times 10^{25} - 4 \times
10^{27}$ W/Hz, derived independently of bin size from the integral LF, is $17\pm11$
Gpc$^{-3}$ to be compared with the value of $40$ Gpc$^{-3}$ in the range $6 \times
10^{24} - 3 \times 10^{27}$ W/Hz for the 1 Jy LF.
\begin{figure}[!h]
\begin{center}
\includegraphics[height=8.cm]{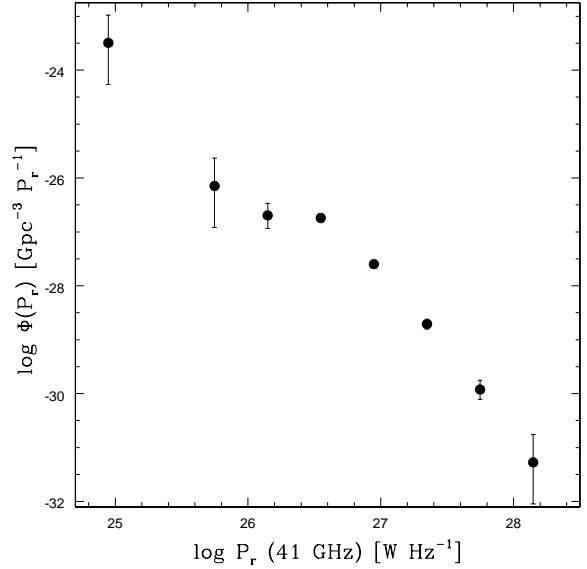}
\end{center}
\caption{The local, de-evolved 41 GHz luminosity function of WMAP FSRQs.
Error bars correspond to $1\sigma$ Poisson errors (\cite{Gehrels86}).} \label{fig.lf_fsrq}
\end{figure}

\subsection{FSRQs}

The local LF of WMAP FSRQs is shown in Fig. \ref{fig.lf_fsrq}.
Based on the $\langle V/V_{\rm max} \rangle$ analysis this has been de-evolved to zero
redshift using $\tau = 0.35$.

We fitted the LF with a single power law $\phi(P_{\rm r}) \propto
P_{\rm r}^{-B_r}$. Varying the binning, the differential slope is
in the range $2.1 < B_r < 2.5$. For a bin size $\Delta \log P_r =
0.4$, which is representative, a weighted least-squares fit yields
$\phi(P_{\rm r}) \propto P_{\rm r}^{-2.3\pm0.1}$ ($\chi^2_{\nu}
\sim 5.1$ for 6 degrees of freedom). A single power law cannot
then reproduce the observed LF (see Fig. \ref{fig.lf_fsrq}). Note
however that the first low-power bin includes a single source (3C
120) and the somewhat large gap between the first and the second
bin. (We note that other authors have classified 3C 120 as a broad
line radio galaxy).

The total number density of FSRQs in the luminosity range $6 \times 10^{24} -
10^{28}$ W/Hz, derived independently of bin size from the integral LF, is $30\pm28$
Gpc$^{-3}$. Excluding the lowest luminosity source one gets $1.6\pm0.4$ Gpc$^{-3}$ in
the luminosity range $7 \times 10^{25} -10^{28}$ W/Hz.
\begin{figure}[!h]
\begin{center}
\includegraphics[height=8.cm]{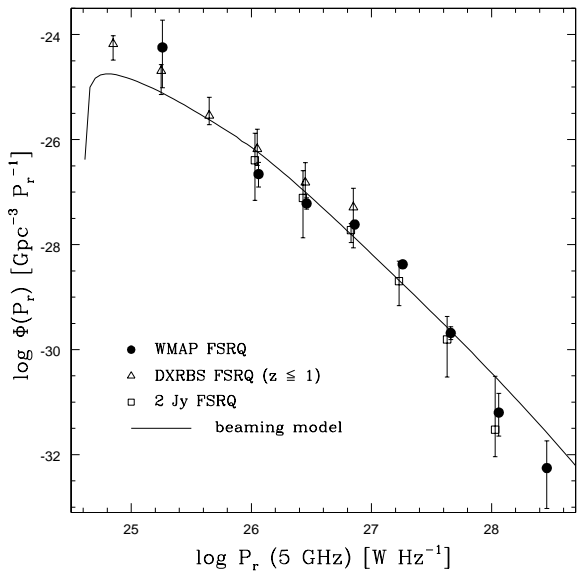}
\end{center}
\caption{The local, de-evolved bivariate 5 GHz radio luminosity function of WMAP FSRQs (filled points) and DXRBS with $z \le 1$ (open triangles) compared to the
predictions of a beaming model based on the 2 Jy luminosity function and evolution (solid
line \cite{Urry95}). The open squares represent the 2 Jy luminosity function. Error bars
represent the sum in quadrature of the $1\sigma$ Poisson errors (\cite{Gehrels86}) and
the variations of the number density associated with a $1\sigma$ change in the
evolutionary parameter $\tau$. For consistency with \cite{Urry95} an empty cosmology with
$H_0 = 50$ km s$^{-1}$ Mpc$^{-1}$, $\Omega_{\rm M} = 0$, and $\Omega_{\rm \Lambda} = 0$
has been adopted. Both beaming model and 2 Jy LF have been converted from 2.7 GHz
assuming $\alpha_{\rm r} = 0$.} \label{fig.lf_fsrq_dxrbs}
\end{figure}

As already done for BL Lacs, we have also derived the local
bivariate LF of FSRQs at 5 GHz for an empty Universe cosmology, to
compare it with previous determinations and with the predictions
of unified schemes. Based on the $\langle V/V_{\rm max} \rangle$
analysis this has been de-evolved using the value $\tau = 0.26$.
The bivariate LF is show in Fig. \ref{fig.lf_fsrq_dxrbs} (filled
points), which also shows the DXRBS (open triangles,
\cite{Pad07a}) and 1 Jy LFs (open squares, \cite{Sti91}), and the
predictions of a beaming model based on the 2 Jy LF and evolution
(solid line, \cite{Urry95}). Given that the DXBRS evolutionary
parameter is epoch dependent and that therefore we cannot simply
de-evolve the global LF to zero redshift, we have restricted
ourselves to sources having $z \le 1$ (see \cite{Pad07a} for
details). We note that 36/52 of the 2 Jy and 8/129 of the DXRBS
FSRQ belong to the WMAP sample.

The WMAP LF is in very good agreement, in the region of overlap,
with the DXRBS and 2 Jy LFs. In particular, the WMAP and 2 Jy LFs
cover roughly the same power range, as expected given the
relatively similar flux limits. The WMAP LF is also in relatively
good agreement with the predictions of unified schemes. Namely,
most bins agree within $1 - 2 \sigma$ with the beaming model, with
the only exception of the two bins around $10^{27}$ W/Hz.

\subsection{Radio galaxies and SSRQs}

\begin{figure}[!h]
\begin{center}
\includegraphics[height=8.cm]{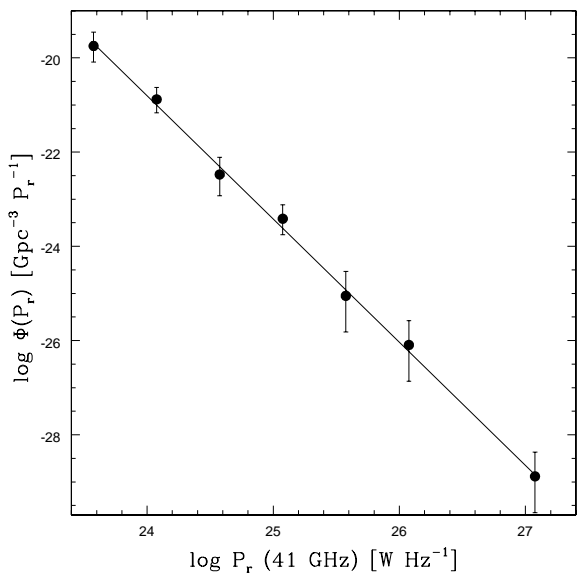}
\end{center}
\caption{The 41 GHz luminosity function of radio galaxies (filled points).
Error bars correspond to $1\sigma$ Poisson errors (\cite{Gehrels86}). The solid
line is a weighted least-squares fit to the data of the form $\phi(P_{\rm r})
\propto P_{\rm r}^{-2.6}$ (see text for details).}
\label{fig.lf_rg}
\end{figure}

The LF of WMAP radio galaxies  is shown in Fig.
\ref{fig.lf_rg}. Based on the $\langle V/V_{\rm max} \rangle$ analysis no evolution has
been assumed. We fitted the LF with a single power law $\phi(P_{\rm r}) \propto P_{\rm
r}^{-B_r}$. Varying the binning, the differential slope is in the range $2.4 < B_r <
2.7$. For a representative bin size $\Delta \log P_r = 0.5$, a weighted least-squares
fit yields $\phi(P_{\rm r}) \propto P_{\rm r}^{-2.6\pm0.1}$ ($\chi^2_{\nu} \sim 0.17$ for
5 degrees of freedom). The total number density of radio galaxies in the luminosity range
$2 \times 10^{23} - 10^{27}$ W/Hz, derived independently of bin size from the integral
LF, is $10,500\pm5,400$ Gpc$^{-3}$.

As regards SSRQs, their sample is too small to derive a
meaningful LF. However, we can say that their total number density in the luminosity
range $5 \times 10^{24} - 4 \times 10^{27}$ W/Hz, derived independently
of bin size from the integral LF, is $91\pm59$ Gpc$^{-3}$.

\section{Discussion and Conclusions}

In this paper we have presented a detailed study of the counterparts of all the
sources of the WMAP 3-year catalogue, which has led to the identification of
309 microwave-selected objects. We have found that only 7 sources (three
planetary nebulae, one Lynds Bright Nebula and three starburst galaxies) are
neither blazars nor non-thermal misdirected AGN (i.e. radio galaxies and
steep-spectrum radio quasars). All the remaining 302 sources have been
classified according to the scheme presented in Sect.2.2. This result fully
confirms the previous findings of \cite{giocol04} that the vast majority of
WMAP detected sources are blazars or radio galaxies.

We have used the WMAP 3-year source catalogue, after correcting their fluxes
for source confusion within the WMAP 41 GHz channel beam, to define a microwave
flux--limited, virtually complete, sample of blazars with $F_{41GHz} > 1$ Jy.
Only 3 out of the 203 objects in this sample still remain unidentified.
This is the first statistically well defined and complete sample of blazars
selected at microwave frequencies.
The microwave frequency range is likely the best region of the electromagnetic spectrum
to pursue statistical studies of blazars since it is least affected by the superposition
of spectral components of different nature like e.g. steep radio emission from the extended
part of the jet, non-nuclear optical emission from the host galaxy or optical/UV and soft
X-ray emission from an accretion disk.

Our complete sample allows us to derive number counts, luminosity function and
cosmological evolution of WMAP AGN.
We found that the LogN-LogS of microwave-selected blazars is steep, $N(>S) \sim
S^{-1.64}$, down to 1 Jy, consistently with that found at cm frequencies (Giommi \&
Colafrancesco 2004, 2005, Giommi et al. 2006).

The bivariate 5 GHz luminosity functions of FSRQs and BL Lacs are
similar to those derived from the DXRBS (for FSRQs and BL Lacs
Padovani et al. 2007) and from the 2--Jy and 1--Jy FSRQs and BL
Lacs surveys, respectively. A beaming model (Urry \& Padovani
1995) adapted to the 1--Jy and 2--Jy blazar surveys is able to
reproduce the observed luminosity function of WMAP blazars down to
$P_r(5GHz) \sim 10^{25}$ W/Hz.\\
The cosmological evolution of these two subclasses of AGN is consistent with the
expectations based on radio/cm measurements, yielding evolution at the $2\sigma$ level
for BL Lacs and a quite strong luminosity
evolution with a time scale $\tau = 0.35^{+0.07}_{-0.05}$ ($\tau =
0.26^{+0.05}_{-0.03}$) for a best-fit $\Lambda$ CDM universe (an empty universe)
reference model.

The luminosity function of radio galaxies can be fit with a power
law shape with slope in the range $2.4 - 2.7$ in the luminosity
interval $P_r(41 GHz) \approx 10^{23.3} - 10^{27}$ W/Hz. As these
sources reach the smallest powers, they have the largest number
density. Therefore, they are bound to be the dominant non-thermal
AGN in the microwave sky at low flux densities. We find here a
value of $\langle V/V_{\rm max} \rangle = 0.61\pm0.07$, consistent
with no evolution. However, this is easily explained by a
combination of small number statistics, as the sample includes
only 15 sources, small redshift range, and a mix of sub-classes. A
look at the radio morphology of our radio galaxies, from both the
literature and the NASA Extragalactic Database (NED), shows in
fact a prevalence of FR I sources (9) followed by FR IIs (4), one
GigaHertz Peaked Spectrum (GPS) and one compact steep-spectrum
object. While FR Is are thought not to be evolving, FR IIs appear
to evolve almost as strongly as radio quasars (see \cite{Urry95}
and references therein). We find a hint of this in our data as
well, since $\langle V/V_{\rm max} \rangle = 0.57\pm0.10$ for FR
Is and $\langle V/V_{\rm max} \rangle = 0.67\pm0.12$ for non-FR
Is.

The present complete sample of microwave-selected non-thermal AGN
will allow us to predict the blazar contribution to the CMB maps
obtainable with coming microwave experiments like {\it Planck}
(see e.g., the {\it Planck} Blue Book) and Olimpo
(\cite{masietal2005}).
The results obtained from our WMAP sample suggest that FSRQs and BL Lacs will
likely dominate the CMB anisotropy spectrum at high microwave fluxes, while the
other non-thermal AGN with lower fluxes but higher space density will provide a relevant
contribution to the CMB anisotropies at lower microwave fluxes, if their LogN-LogS
remains steep, as suggested by previous analyses at radio frequency.

The large number of faint non-blazar AGN will further provide a substantial
contribution to the diffuse light (effective temperature) of the CMB sky observable with
up-coming all-sky surveys. The additional contribution of microwave faint starburst
galaxies (which are only marginally present in the high-flux threshold WMAP sample) will
likely provide a relevant additional contribution to the unresolved CMB foreground light.
We will present a detailed analysis of the contribution of different classes of non-thermal AGN to
the cosmic background radiation field in a forthcoming paper (Padovani et al. 2009, in
preparation).\\
Detailed considerations about the contribution of the polarized non-thermal AGN to the CMB
polarization anisotropy spectrum will also be presented elsewhere (Colafrancesco et al.
2009, in preparation).

All the non-thermal AGNs (blazars, SSRQs and radio galaxies) in our sample are bright, well known
sources that have been detected in several energy bands (from radio to X-rays) and are
also expected to be detected in the gamma-ray band by the AGILE and {\it Fermi}
experiments (see below). In this respect, almost 50 WMAP sources have already a counterpart in the
third EGRET catalogue of gamma-ray sources (\cite{Hartman1999}). In particular, AGILE
is expected to detect a similar number of blazars and will monitor their duty cycle.

Finally, we note that the ratio of the number of FSRQs and BL Lacs
at radio or microwave frequencies is similar to that found at
gamma-ray energies. In fact, this ratio is $\approx 6$ in our sample (see Table \ref{tab:evol}),
$\approx 8$ in the 5GHz DXRBS survey \cite{Pad07a}, and $\approx 3-5$ for
the EGRET blazars \cite{SE04}).

Since the radio fluxes of EGRET detected blazars are similar to
those of sources detected in the WMAP survey, blazar count ratios
can be in agreement only with scenarios where the production of
gamma-rays (that is intimately related to radio/microwave photons
in Synchrotron - Inverse Compton models) is symmetric with respect
to the two blazar types or slightly favour BL Lacs. Mechanisms
like Synchrotron-Inverse Compton with External components (see,
e.g., Sikora \& Madejski 2001 for a review), which strongly favour
FSRQs cannot be therefore very common otherwise blazars of this
type would be much more abundant than BL Lacs at gamma-ray
frequencies. The recent {\it Fermi} results appear to confirm
this. In fact, \cite{abdo09} find an FSRQ/BL Lac ratio of 1.3.
This relatively low ratio is due to a strong selection effect,
which picks up radio-faint BL Lacs with the synchrotron peak at
high energies characterized by a flat gamma-ray spectrum. By
comparing FSRQs and BL Lacs with similar synchrotron peaks
(\cite{gio09}), one obtains a ratio $\approx 4$, not too different
from that of WMAP and EGRET. In any case, even {\it Fermi} sees no
evidence of enhanced gamma-ray emission from FSRQs as compared to
that from BL Lacs.

We can conclude that the microwave and gamma-ray energy bands are best suited to study non-thermal
AGN as their emission at these frequencies is largely dominated by non-thermal
radiation. The combined study of non-thermal AGN in the microwave and in the gamma-ray energy
bands will probably offer a unique opportunity to understand many of the physical details of
this class of BH-dominated cosmic structures.

\begin{acknowledgements}
We are grateful to F. Verrecchia for his technical support in the
analysis of the WMAP source confusion. We acknowledge the use of
data and software facilities from the ASI Science Data Center
(ASDC), managed by the Italian Space Agency (ASI). This work
benefits from ASI grant I/024/05/1. Part of this work is based on
archival data and bibliographic information obtained from the NASA
Extragalactic Database (NED) and from the Astrophysics Data System
(ADS).
\end{acknowledgements}

 \onecolumn
 \scriptsize
\begin{landscape}
\begin{longtable}{lcccccccccc}
 \caption{WMAP3 catalogue: basic data, identifications and inclusion in the flux limited sample} \\
 \hline
 \multicolumn{1}{c}{Source name} & \multicolumn{1}{c}{WMAP ID} &
 \multicolumn{1}{c}{Common name} & \multicolumn{1}{r}{Ra(J2000.0)} &
 \multicolumn{1}{r}{Dec(J2000.0)} & \multicolumn{1}{c}{Class} &
 \multicolumn{1}{c}{Redshift} & \multicolumn{1}{c}{Radio flux} &
 \multicolumn{1}{c}{WMAP flux} & \multicolumn{1}{c}{Corrected} &
 \multicolumn{1}{c}{In Complete} \\
 \multicolumn{1}{c}{IAU Convention} &  \multicolumn{1}{c}{1year catalogue}
 & & & & &     &  density 5 GHz & density 41 GHz  & 41 GHz    &   \multicolumn{1}{c}{Sample} \\
 & & & & & &     &  (mJy)   & (mJy)      & (mJy)            &   \\
 \hline
\endfirsthead
\caption{WMAP3 catalogue: basic data, identifications and inclusion in the flux limited
sample (continued)}\\
 \hline
 \multicolumn{1}{c}{Source name} & \multicolumn{1}{c}{WMAP ID} &
 \multicolumn{1}{c}{Common name} & \multicolumn{1}{r}{Ra(J2000.0)} &
 \multicolumn{1}{r}{Dec(J2000.0)} & \multicolumn{1}{c}{Class} &
 \multicolumn{1}{c}{Redshift} & \multicolumn{1}{c}{Radio Flux} &
 \multicolumn{1}{c}{WMAP flux} & \multicolumn{1}{c}{Corrected} &
  \multicolumn{1}{c}{In Complete} \\
 \multicolumn{1}{c}{IAU Convention}  &  \multicolumn{1}{c}{1year catalogue}
 & & & & &     &  density 5 GHz & density 41 GHz  & 41 GHz    &   \multicolumn{1}{c}{Sample}    \\
 & & & & & &     &  (mJy)   & (mJy)      & (mJy)     &      \\
 \hline
 \endhead
 \hline
 \endfoot

WMAP3 J0006$-$0622&WMAP 60&PKS0003-066     &00 06 13.8&$-$06 23
35.0&FSRQ&0.347&2463&2300&2300&yes\\ WMAP3
J0012$-$3953&WMAP 202&PKS0010-401     &00 12 59.8&$-$39 54
24.9&BL Lac&0&921&1000&1000&yes\\ WMAP3
J0019+2019&---&PKS0017+200     &00 19 37.7&+20 21 45.0&BL Lac&0&710&900&900&no\\ WMAP3 J0019+2602&---&4C25.01         &00
19 39.1&+26 02 44.9&FSRQ&0.284&435&900&900&no\\ WMAP3
J0025$-$2604&---&PKS0023-26      &00 25 49.0&$-$26 02 12.9&Radio Galaxy&0.322&3745&0&0&no\\ WMAP3
J0026$-$3511&---&PMNJ0026-3512   &00 26 16.3&$-$35 12
48.9&Unidentified &0&121&1200&1200&yes\\ WMAP3
J0029+0555&---&PKS0027+056     &00 29 45.7&+05 54 41.0&FSRQ&1.317&348&1200&1100&yes\\ WMAP3 J0042+5208&---&3C020
&00 43 08.7&+52 03 33.9&Radio Galaxy&0.174&4124&0&0&no\\
WMAP3 J0047$-$2515&WMAP 62&NGC253          &00 47 33.1&$-$25 17
17.0&Starburst galaxy&0.001&2433&1100&1100&yes\\ WMAP3
J0049$-$5740&WMAP 179&PKS0047-579     &00 49 59.4&$-$57 38
26.9&FSRQ&1.797&1338&1000&1000&yes\\ WMAP3
J0051$-$0927&WMAP 77&PKS0048-09      &00 50 41.2&$-$09 29
04.9&BL Lac&0.537&931&800&800&no\\ WMAP3
J0050$-$0647&---&PKS0048-71      &00 51 08.2&$-$06 50 02.0&FSRQ&1.975&841&800&800&no\\ WMAP3 J0050$-$4223&---&PKS0048-427
&00 51 09.4&$-$42 26 33.0&FSRQ&1.749&926&900&900&no\\
WMAP3 J0106$-$4035&WMAP 171&PKS0104-408     &01 06 45.1&$-$40 34
19.9&FSRQ&0.584&2584&1800&1800&yes\\ WMAP3
J0108+0135&WMAP 81&PKS0106+01      &01 08 37.6&+01 34 59.9&FSRQ&2.099&2078&2300&2300&yes\\ WMAP3 J0108+1320&WMAP 79&3C033
&01 08 52.7&+13 19 13.0&Radio Galaxy&0.059&3691&1000&900&no\\ WMAP3
J0115$-$0126&---&PKS0112-017     &01 15 17.1&$-$01 27 05.0&FSRQ&1.365&1540&800&800&no\\ WMAP3 J0116$-$1137&---&PKS0113-118
&01 16 12.4&$-$11 36 15.0&Blazar
Unknown-type&0.67&1488&1200&1200&yes\\ WMAP3
J0121+1150&---&PKS0119+11      &01 21 41.5&+11 49 50.0&FSRQ&0.57&1126&1200&1200&yes\\ WMAP3 J0125$-$0010&WMAP
86&PKS0122-00      &01 25 28.8&$-$00 05 56.0&FSRQ&1.076&1234&1300&1300&yes\\ WMAP3 J0132$-$1653&WMAP
97&PKS0130-17      &01 32 43.3&$-$16 54 47.9&FSRQ&1.02&746&1900&1900&yes\\ WMAP3 J0137+4752&WMAP 80&S40133+47
&01 36 58.4&+47 51 29.0&FSRQ&0.859&2016&4200&4200&no\\
WMAP3 J0137$-$2428&---&PKS0135-247     &01 37 38.2&$-$24 30
54.0&FSRQ&0.837&956&1700&1700&yes\\ WMAP3
J0152+2208&---&PKS0149+21      &01 52 17.9&+22 07 06.9&FSRQ&1.32&1073&1200&1200&yes\\ WMAP3 J0204+1513&WMAP 92&4C15.05
&02 04 50.2&+15 14 11.0&FSRQ&0.405&3073&1500&1500&yes\\
WMAP3 J0204+3213&WMAP 85&1JY0202+319     &02 05 04.9&+32 12
29.9&FSRQ&1.446&934&1200&1200&yes\\ WMAP3
J0210$-$5100&WMAP 158&PKS0208-512     &02 10 46.0&$-$51 01
01.9&FSRQ&0.999&3198&2800&2800&yes\\ WMAP3
J0220+3558&---&1Jy0218+357     &02 21 05.4&+35 56 13.9&Blazar
Unknown-type&0.685&1480&1000&1000&yes\\ WMAP3 J0223+4304&WMAP
84&3C66B           &02 23 11.3&+42 59 30.9&Radio Galaxy&0.021&1866&1400&1200&yes\\ WMAP3 J0222$-$3441&WMAP
137&PKS0220-349     &02 22 56.4&$-$34 41 29.0&FSRQ&1.49&911&0&0&no\\ WMAP3 J0231+1318&---&4C13.14         &02
31 45.7&+13 22 54.9&FSRQ&2.059&2608&1300&1300&yes\\ WMAP3
J0237+2848&WMAP 93&4C28.07         &02 37 52.3&+28 48 08.9&FSRQ&1.213&2794&3500&3500&yes\\ WMAP3 J0238+1636&---&PKS0235+164
&02 38 38.8&+16 36 59.0&BL Lac&0.94&1935&1400&1400&yes\\
WMAP3 J0253$-$5441&WMAP 155&PKS0252-549     &02 53 29.1&$-$54 41
50.9&FSRQ&0.539&1193&2800&2800&yes\\ WMAP3
J0303$-$6212&WMAP 162&PKS0302-623     &03 03 50.5&$-$62 11
25.0&Blazar Unknown-type&0&1862&1400&1400&yes\\ WMAP3
J0308+0404&WMAP 102&NGC1218         &03 08 26.2&+04 06 39.9&Radio Galaxy&0.028&3773&1400&1400&yes\\ WMAP3
J0309$-$6103&WMAP 160&PKS0308-611     &03 09 55.9&$-$60 58
38.9&Blazar Unknown-type&0&1103&900&900&no\\ WMAP3
J0312$-$7647&WMAP 174&PKS0312-77      &03 11 55.1&$-$76 51
51.0&FSRQ&0.223&694&1100&1100&yes\\ WMAP3 J0319+4131&WMAP
94&NGC1275         &03 19 48.1&+41 30 42.0&Radio Galaxy&0.017&46894&7000&7000&no\\ WMAP3 J0322$-$3711&WMAP
138&NGC1316         &03 22 41.7&$-$37 12 38.9&Radio Galaxy&0.006&71133&10500&10500&yes\\ WMAP3
J0325+2224&---&GB6J0325+2223   &03 25 36.7&+22 23 59.9&FSRQ&2.06&817&1100&1100&yes\\ WMAP3 J0329$-$2353&WMAP
123&PKS0327-241     &03 29 54.0&$-$23 57 09.0&FSRQ&0.895&866&1200&1200&yes\\ WMAP3 J0334$-$4007&WMAP
146&PKS0332-403     &03 34 13.6&$-$40 08 25.0&BL Lac&0&1331&1600&1600&yes\\ WMAP3 J0336$-$1258&---&PKS0334-131
&03 36 34.9&$-$13 02 03.9&FSRQ&1.303&534&1200&1200&yes\\
WMAP3 J0339$-$0144&WMAP 106&PKS0336-019     &03 39 30.8&$-$01 46
36.0&FSRQ&0.852&3014&2700&2500&yes\\ WMAP3
J0340$-$2121&---&PKS0338-214     &03 40 35.5&$-$21 19 31.0&BL Lac&0.223&933&1200&1200&yes\\ WMAP3 J0348$-$2747&WMAP
129&PKS0346-279     &03 48 38.0&$-$27 48 12.9&FSRQ&0.991&1444&800&800&no\\ WMAP3 J0358+1029&---&3C098
&03 58 54.4&+10 26 03.0&Radio Galaxy&0.03&3132&0&0&no\\
WMAP3 J0403$-$3604&WMAP 136&PKS0402-362     &04 03 53.7&$-$36 05
02.0&FSRQ&1.417&1851&4500&4500&yes\\ WMAP3
J0405$-$1305&WMAP 114&PKS0403-13      &04 05 34.0&$-$13 08
12.9&FSRQ&0.57&2748&1800&1800&yes\\ WMAP3 J0411+7655&WMAP
82&4C76.03         &04 10 45.5&+76 56 44.9&Radio Galaxy/GPS&0.598&5620&800&800&no\\ WMAP3 J0407$-$3824&WMAP
141&PKS0405-385     &04 06 58.9&$-$38 26 27.9&Blazar
Unknown-type&1.285&830&1000&1000&yes\\ WMAP3
J0408$-$7503&---&PKS0410-75      &04 08 48.4&$-$75 07 18.9&Radio Galaxy&0.693&4569&600&600&no\\ WMAP3
J0423+0219&---&PKS0420+022     &04 22 52.2&+02 19 27.0&FSRQ&0&1222&0&0&no\\ WMAP3 J0423$-$0120&WMAP 110&PKS0420-01
&04 23 15.7&$-$01 20 33.0&FSRQ&0.914&4357&10100&10100&yes\\ WMAP3 J0424+0036&WMAP
109&PKS0422+004     &04 24 46.7&+00 36 06.0&BL Lac&0.31&583&1800&1700&yes\\ WMAP3 J0424$-$3757&WMAP
140&PKS0422-380     &04 24 42.1&$-$37 56 20.0&FSRQ&0.782&1706&1300&1300&yes\\ WMAP3
J0428$-$3757&---&PKS0426-380     &04 28 40.3&$-$37 56 20.0&BL Lac&1.11&1202&1300&1300&yes\\ WMAP3 J0433+0521&WMAP 108&3C120
&04 33 10.9&+05 21 15.0&FSRQ&0.033&5189&2600&2600&yes\\
WMAP3 J0440$-$4333&WMAP 147&PKS0438-43      &04 40 17.1&$-$43 33
09.0&FSRQ&2.863&3933&2800&2800&yes\\ WMAP3
J0442$-$0018&---&PKS0440-00      &04 42 38.5&$-$00 17 43.0&FSRQ&0.844&1084&1200&1200&yes\\ WMAP3 J0453$-$2806&WMAP
131&PKS0451-28      &04 53 14.5&$-$28 07 36.9&FSRQ&2.559&2172&0&0&no\\ WMAP3 J0449$-$8100&WMAP 175&PKS0454-81
&04 50 05.3&$-$81 01 01.9&FSRQ&0.444&1357&1700&1700&yes\\
WMAP3 J0455$-$4617&WMAP 151&PKS0454-46      &04 55 50.7&$-$46 15
59.0&FSRQ&0.853&1653&4000&4000&yes\\ WMAP3
J0457$-$2322&WMAP 128&PKS0454-234     &04 57 03.1&$-$23 24
51.9&FSRQ&1.003&1863&2600&2600&yes\\ WMAP3
J0501$-$0158&---&4C-02.19        &05 01 12.7&$-$01 59 13.9&FSRQ&2.286&3317&1300&1200&yes\\ WMAP3 J0506$-$6108&WMAP
154&PKS0506-61      &05 06 43.9&$-$61 09 41.0&FSRQ&1.093&1211&1700&1500&yes\\ WMAP3 J0513$-$2155&WMAP
127&PKS0511-220     &05 13 49.0&$-$21 59 16.0&FSRQ&1.296&865&1000&1000&yes\\ WMAP3
J0513$-$2015&---&PMNJ0514-2029   &05 14 17.2&$-$20 29
21.0&Unidentified &0&431&800&700&no\\ WMAP3
J0515$-$4558&---&PKS0514-459     &05 15 45.3&$-$45 46 44.0&FSRQ&0.194&0&900&400&no\\ WMAP3 J0519$-$0540&WMAP 116&
&05 19 14.5&$-$05 40 41.0&LBN Nebula&0&0&1800&1800&yes\\ WMAP3
J0519$-$4546&WMAP 150&PICTORA         &05 19 49.6&$-$45 46
44.0&SSRQ&0.035&15827&4200&4200&yes\\ WMAP3
J0523$-$3627&WMAP 139&PKS0521-36      &05 22 57.8&$-$36 27
30.9&BL Lac&0.055&8180&3400&3400&yes\\ WMAP3
J0525$-$4826&---&PKS0524-485     &05 26 16.6&$-$48 30
35.9&Unidentified &0&425&1500&1500&yes\\ WMAP3 J0527$-$1240&WMAP
122&IC0418          &05 27 28.2&$-$12 41 49.9&Planetary
Nebula&0&1716&1400&1400&yes\\ WMAP3 J0538$-$4405&WMAP
148&PKS0537-441     &05 38 50.2&$-$44 05 08.9&BL Lac&0.894&4805&5700&5700&yes\\ WMAP3 J0542+4951&WMAP 95&3C147.0
&05 42 36.1&+49 51 06.9&SSRQ&0.545&7264&1100&1100&no\\ WMAP3 J0540$-$5416&WMAP
152&PKS0539-543     &05 40 45.7&$-$54 18 21.9&FSRQ&1.185&373&1800&1800&yes\\ WMAP3 J0550$-$5732&WMAP
153&PKS0549-575     &05 50 09.4&$-$57 32 23.9&FSRQ&2.001&355&1200&1200&yes\\ WMAP3 J0555+3944&WMAP
100&B20552+39A      &05 55 30.7&+39 48 48.9&FSRQ&2.365&5524&1100&1100&no\\ WMAP3
J0600$-$4528&---&PKS0557-454     &05 59 11.5&$-$45 29 39.9&FSRQ&0.687&407&900&900&no\\ WMAP3 J0607+6722&WMAP 91&S40602+673
&06 07 52.0&+67 20 54.9&FSRQ&1.97&581&700&700&no\\ WMAP3
J0608$-$2219&---&PKS0606-223     &06 08 59.7&$-$22 20 21.0&FSRQ&1.926&626&1000&1000&yes\\ WMAP3 J0609$-$1541&WMAP
126&PKS0607-15      &06 09 40.9&$-$15 42 41.0&FSRQ&0.324&2473&3400&3400&yes\\ WMAP3
J0623$-$6436&---&IRASL06229-643  &06 23 07.7&$-$64 36 20.9&FSRQ&0.129&531&900&900&no\\ WMAP3 J0629$-$1957&WMAP
130&PKS0627-199     &06 29 23.7&$-$19 59 20.0&BL Lac&0&784&1600&1600&no\\ WMAP3 J0633$-$2217&WMAP
135&PMNJ0633-2223   &06 33 26.7&$-$22 23 21.0&FSRQ&1.508&885&900&900&no\\ WMAP3 J0639+7326&WMAP 87&S50633+734
&06 39 21.8&+73 24 57.9&FSRQ&1.85&711&1100&1100&yes\\
WMAP3 J0636$-$2032&WMAP 134&PKS0634-20      &06 36 32.2&$-$20 34
53.0&Radio Galaxy&0.055&1326&900&900&no\\ WMAP3
J0635$-$7517&WMAP 167&PKS0637-75      &06 35 46.4&$-$75 16
17.0&FSRQ&0.653&6398&4000&4000&yes\\ WMAP3
J0646+4449&WMAP 99&S40642+449      &06 46 31.9&+44 51 15.9&FSRQ/GPS&3.396&1220&2200&2200&yes\\ WMAP3 J0721+7122&---&S50716+714
&07 21 53.4&+71 20 35.9&BL Lac&0.3&859&2200&2200&yes\\
WMAP3 J0721+0401&---&GB6J0721+0406   &07 21 23.8&+04 06
46.0&Unidentified &0&533&700&700&no\\ WMAP3 J0727+6748&---&3C179
&07 28 11.2&+67 48 47.0&SSRQ&0.846&912&700&700&no\\ WMAP3 J0725$-$0049&---&PKS0723-008
&07 25 50.5&$-$00 54 56.0&Blazar
Unknown-type&0.128&1377&1200&1200&no\\ WMAP3
J0734+5021&---&GB6J0733+5022   &07 33 52.4&+50 22 09.0&FSRQ&0.72&991&1200&1200&yes\\ WMAP3 J0738+1744&WMAP
113&PKS0735+17      &07 38 07.3&+17 42 19.0&BL Lac&0.42&1812&1300&1200&yes\\ WMAP3 J0739+0136&WMAP
124&PKS0736+01      &07 39 18.0&+01 37 04.0&FSRQ&0.19&1781&2500&2500&no\\ WMAP3 J0741+3111&WMAP 107&OI363
&07 41 10.6&+31 11 59.9&FSRQ&0.63&2623&900&900&no\\ WMAP3
J0745+1012&WMAP 118&PKS0742+10      &07 45 33.0&+10 11 12.9&FSRQ/GPS&2.624&3704&800&800&no\\ WMAP3 J0746$-$0045&---&4C-00.28
&07 45 54.1&$-$00 44 17.0&FSRQ/GPS&0.994&1975&800&800&no\\
WMAP3 J0743$-$6727&WMAP 161&PKS0743-67      &07 43 31.6&$-$67 26
25.0&FSRQ&1.51&2235&700&600&no\\ WMAP3 J0750+1230&WMAP
117&PKS0748+126     &07 50 52.0&+12 31 05.0&FSRQ&0.889&1238&2600&2600&yes\\ WMAP3 J0753+5354&---&4C54.15
&07 53 01.3&+53 52 59.0&BL Lac&0&964&1000&1000&yes\\ WMAP3
J0757+0957&WMAP 120&PKS0754+100     &07 57 06.6&+09 56 35.0&BL Lac&0.266&1037&1600&1600&yes\\ WMAP3 J0808$-$0750&WMAP
133&PKS0805-07      &08 08 15.4&$-$07 51 10.0&FSRQ&1.837&1599&1400&1400&no\\ WMAP3 J0813+4817&---&3C196
&08 13 35.9&+48 13 00.9&SSRQ&0.871&4504&0&0&no\\
WMAP3 J0816$-$2425&WMAP 145&PMNJ0816-2421   &08 16 40.3&$-$24 21
05.0&Unidentified &0&270&1200&1200&no\\ WMAP3
J0824+3914&---&4C+39.23        &08 24 55.5&+39 16 41.9&FSRQ&1.216&1031&1000&900&no\\ WMAP3 J0825+0311&WMAP
125&PKS0823+033     &08 25 50.2&+03 09 24.0&BL Lac&0.506&2064&1500&1500&yes\\ WMAP3 J0830+2410&WMAP
112&B20827+24       &08 30 52.0&+24 10 59.9&FSRQ&0.94&886&1600&1600&yes\\ WMAP3 J0838+5820&---&S40833+58
&08 37 22.3&+58 25 00.9&FSRQ&2.101&717&1000&1000&yes\\
WMAP3 J0836$-$2015&WMAP 144&PKS0834-20      &08 36 39.1&$-$20 16
59.0&FSRQ&2.752&1712&2300&2300&no\\ WMAP3 J0841+7053&WMAP
89&4C71.07         &08 41 24.2&+70 53 41.9&FSRQ&2.172&2342&1800&1800&yes\\ WMAP3 J0840+1312&WMAP
121&3C207.0         &08 40 47.6&+13 12 23.0&SSRQ&0.681&1244&1800&1800&yes\\ WMAP3 J0854+2005&WMAP
115&PKS0851+202     &08 54 48.7&+20 06 29.9&BL Lac&0.306&2908&4200&4200&yes\\ WMAP3
J0902$-$1414&---&PKS0859-14      &09 02 16.8&$-$14 15 30.9&FSRQ&1.333&2022&1300&1300&yes\\ WMAP3
J0907$-$2017&---&PMNJ0907-2026   &09 07 54.1&$-$20 26
51.0&Unidentified &0&417&0&0&no\\ WMAP3 J0909+4252&---&3C216
&09 09 33.4&+42 53 47.0&Radio Galaxy&0.67&1561&1200&1200&yes\\ WMAP3 J0909+0118&WMAP
132&4C01.24         &09 09 10.0&+01 21 36.0&FSRQ&1.024&888&1800&1800&yes\\ WMAP3 J0914+0249&---&PKS0912+029
&09 14 37.9&+02 45 59.0&FSRQ&0.427&842&1400&1400&yes\\
WMAP3 J0918$-$1203&WMAP 143&HYDRAA          &09 18 05.6&$-$12 05
44.0&Radio Galaxy&0.055&13982&1000&1000&yes\\ WMAP3
J0921+6216&---&S40917+62       &09 21 36.1&+62 15 51.9&FSRQ&1.446&1226&900&900&no\\ WMAP3 J0920+4440&---&S40917+44
&09 20 58.3&+44 41 53.9&FSRQ&2.19&1085&1300&1300&yes\\
WMAP3 J0921$-$2619&---&PKS0919-260     &09 21 29.4&$-$26 18
42.9&FSRQ&2.3&2333&1200&1200&yes\\ WMAP3 J0927+3901&WMAP
105&4C39.25         &09 27 02.8&+39 02 21.0&FSRQ&0.695&6913&5200&5200&yes\\ WMAP3 J0948+4037&WMAP
104&4C40.24         &09 48 55.3&+40 39 45.0&FSRQ&1.249&1801&1400&1400&yes\\ WMAP3 J0955+6936&WMAP 88&M82
&09 55 51.7&+69 40 45.0&Starburst
galaxy&0.001&3796&1100&1100&yes\\ WMAP3 J0957+5526&---&4C55.17
&09 57 38.1&+55 22 58.0&FSRQ&0.896&2015&1000&1000&yes\\
WMAP3 J0958+4721&WMAP 98&S40955+476      &09 58 19.6&+47 25
08.0&FSRQ&1.882&1005&1400&1400&yes\\ WMAP3
J1014+2304&WMAP 119&4C23.24         &10 14 47.1&+23 01 15.9&FSRQ&0.565&1093&900&900&no\\ WMAP3 J1018$-$3129&---&PKS1016-311
&10 18 28.6&$-$31 23 53.9&FSRQ&0.794&688&0&0&no\\ WMAP3
J1032+4118&WMAP 103&S41030+415      &10 33 03.7&+41 16 05.9&FSRQ&1.117&439&900&900&no\\ WMAP3 J1037$-$2934&---&PKS1034-293
&10 37 16.0&$-$29 34 02.9&Blazar
Unknown-type&0.312&1477&1100&1100&yes\\ WMAP3 J1038+0511&WMAP
142&PKS1036+054     &10 38 46.7&+05 12 29.0&FSRQ&0.473&704&1500&1500&yes\\ WMAP3 J1041+0611&---&4C06.41
&10 41 17.0&+06 10 17.0&FSRQ&1.27&1518&1400&1400&yes\\
WMAP3 J1041$-$4740&WMAP 163&PMNJ1041-4740   &10 41 44.2&$-$47 40
00.0&Unidentified &0&2480&300&300&no\\ WMAP3 J1047+7143&WMAP
83&S51044+719      &10 48 27.6&+71 43 35.0&FSRQ&1.15&1900&800&800&no\\ WMAP3 J1047$-$1911&---&PKS1045-18
&10 48 06.5&$-$19 09 36.0&FSRQ&0.595&879&800&800&no\\
WMAP3 J1058+0134&WMAP 149&4C01.28         &10 58 29.5&+01 33
59.0&Blazar Unknown-type&0.89&3263&4500&4500&yes\\ WMAP3
J1059$-$8003&WMAP 176&PKS1057-79      &10 58 43.3&$-$80 03
53.9&Blazar Unknown-type&0&2130&2200&2200&yes\\ WMAP3
J1102$-$4400&---&PKS1059-438     &11 02 04.8&$-$44 04 22.0&FSRQ&0.763&448&0&0&no\\ WMAP3 J1107$-$4446&WMAP 166&PKS1104-445
&11 07 08.5&$-$44 49 08.0&FSRQ&1.598&3325&1200&1200&no\\
WMAP3 J1117$-$4635&---&PKS1116-46      &11 18 26.8&$-$46 34
14.9&FSRQ&0.713&1872&600&600&no\\ WMAP3
J1118+1238&---&4C12.39         &11 18 57.3&+12 34 41.9&FSRQ&2.129&1820&1100&1100&yes\\ WMAP3 J1127$-$1858&WMAP
159&PKS1124-186     &11 27 04.3&$-$18 57 16.9&FSRQ&1.05&1624&1300&1300&yes\\ WMAP3 J1130$-$1452&WMAP
157&PKS1127-145     &11 30 07.0&$-$14 49 27.0&FSRQ/GPS&1.184&4209&1500&1500&yes\\ WMAP3 J1130+3813&WMAP
101&S41128+385      &11 30 53.2&+38 15 18.0&FSRQ&1.741&769&1100&1100&yes\\ WMAP3 J1146$-$4838&---&PKS1143-48
&11 45 30.7&$-$48 36 11.9&Radio Galaxy&0.33&1376&0&0&no\\ WMAP3 J1146+4001&---&S41144+402
&11 46 58.2&+39 58 33.9&FSRQ&1.088&836&1100&1100&yes\\
WMAP3 J1147$-$3811&WMAP 169&PKS1144-379     &11 47 01.3&$-$38 12
11.0&BL Lac&1.048&1825&1900&1700&yes\\ WMAP3
J1149$-$7932&---&PMNJ1147-7936   &11 46 50.5&$-$79 37
09.0&Unidentified &0&0&700&700&no\\ WMAP3 J1154+8104&WMAP
78&S51150+81       &11 53 12.4&+80 58 28.9&FSRQ&1.25&1180&0&0&no\\ WMAP3 J1153+4931&WMAP 90&4C49.22
&11 53 24.4&+49 31 09.0&FSRQ&0.334&717&2200&1600&yes\\
WMAP3 J1159+2914&WMAP 111&4C29.45         &11 59 31.7&+29 14
44.0&FSRQ&0.729&1461&2000&2000&yes\\ WMAP3
J1203+4806&---&GB6J1203+4803   &12 03 29.8&+48 03 12.9&FSRQ&0.817&164&700&700&no\\ WMAP3 J1208$-$2405&WMAP
172&PKSB1206-238    &12 09 02.3&$-$24 06 20.9&FSRQ&1.299&1073&900&900&no\\ WMAP3 J1215$-$1730&WMAP
173&PKS1213-17      &12 15 46.7&$-$17 31 45.9&Blazar
Unknown-type&0&1744&1200&1200&yes\\ WMAP3 J1218+4832&---&S41216+48
&12 19 06.4&+48 29 56.0&FSRQ&1.076&638&800&800&no\\ WMAP3
J1219+0548&WMAP 164&NGC4261         &12 19 23.2&+05 49 31.0&Radio Galaxy&0.007&9064&2000&2000&yes\\ WMAP3
J1222$-$8307&WMAP 178&PKS1221-82      &12 24 54.4&$-$83 13
09.9&Blazar Unknown-type&0&797&900&900&no\\ WMAP3
J1227+1124&---&GB6J1228+1124   &12 28 29.1&+11 24
52.9&Unidentified &0&0&500&500&no\\ WMAP3 J1229+0203&WMAP
170&3C273           &12 29 06.7&+02 03 07.9&FSRQ&0.158&36923&15700&15700&yes\\ WMAP3 J1230+1223&WMAP 165&M87
&12 30 49.3&+12 23 27.9&Radio Galaxy&0.004&59027&13100&13100&yes\\ WMAP3
J1231+1351&---&GB6J1231+1344   &12 31 42.9&+13 44 48.0&No clear
radio counterpart&0.115&0&0&0&no\\ WMAP3 J1246$-$2547&WMAP
177&PKS1244-255     &12 46 46.6&$-$25 47 48.0&FSRQ&0.633&2317&1800&1800&yes\\ WMAP3 J1256$-$0547&WMAP
181&3C279           &12 56 10.9&$-$05 47 21.0&FSRQ&0.536&11192&18000&18000&yes\\ WMAP3 J1258$-$3158&WMAP
180&PKS1255-316     &12 57 58.9&$-$31 55 14.9&FSRQ&1.924&1410&1300&1300&yes\\ WMAP3
J1258$-$2223&---&PKS1256-220     &12 58 54.4&$-$22 19 32.0&FSRQ&1.306&978&1000&1000&yes\\ WMAP3 J1302+4856&---&
&13 02 49.3&+48 56 35.9&No clear radio
counterpart&0&0&900&900&no\\ WMAP3 J1305$-$4928&---&NGC4945
&13 05 27.4&$-$49 28 04.0&FIR Starburst galaxy&0.002&3055&0&0&no\\
WMAP3 J1309+1155&---&MC21307+12      &13 09 33.9&+11 54
24.9&BL Lac&0&1350&1000&1000&yes\\ WMAP3 J1310+3221&WMAP
52&1Jy1308+326     &13 10 28.6&+32 20 43.0&BL Lac&0.996&1447&2700&2000&yes\\ WMAP3 J1316$-$3337&WMAP
182&PKS1313-333     &13 16 07.9&$-$33 38 58.9&FSRQ&1.21&1093&1700&1700&yes\\ WMAP3
J1324$-$1046&---&PKSB1321-105    &13 24 25.7&$-$10 49 23.0&FSRQ&0.872&857&1000&1000&yes\\ WMAP3
J1327+2212&---&GB6J1327+2210   &13 27 00.7&+22 10 50.0&FSRQ&1.4&647&800&800&no\\ WMAP3 J1329+3201&WMAP 40&GB6J1329+3154
&13 29 52.7&+31 54 11.0&BL Lac&0&518&800&800&no\\ WMAP3
J1330+2504&---&3C287           &13 30 37.6&+25 09 10.0&SSRQ&1.055&3289&1100&1100&yes\\ WMAP3 J1331+3031&WMAP
26&3C286           &13 31 08.2&+30 30 32.0&SSRQ&0.849&6350&1400&1400&yes\\ WMAP3 J1332+0200&---&3C287.1
&13 32 53.2&+02 00 45.0&SSRQ&0.216&1369&1100&1100&yes\\ WMAP3
J1333+2723&---&GB6J1333+2725   &13 33 07.3&+27 25 18.0&FSRQ&2.126&318&900&900&no\\ WMAP3 J1336$-$3358&WMAP 185&IC4296
&13 36 43.0&$-$33 58 40.0&Radio Galaxy&0.012&1886&1100&1100&yes\\ WMAP3 J1337$-$1257&WMAP
188&PKS1335-127     &13 37 39.7&$-$12 57 24.0&FSRQ&0.539&2838&6400&6400&yes\\ WMAP3 J1343+6601&---&S41342+663
&13 43 45.9&+66 02 26.0&FSRQ&0.776&545&600&100&no\\ WMAP3
J1354$-$1041&WMAP 197&PKS1352-104     &13 54 46.9&$-$10 41
02.0&FSRQ&0.332&686&1600&1600&yes\\ WMAP3
J1357$-$1529&---&PKS1354-152     &13 57 11.2&$-$15 27 29.0&FSRQ&1.89&841&1000&1000&yes\\ WMAP3 J1356+1919&WMAP 4&4C19.44
&13 57 04.3&+19 19 06.9&FSRQ&0.72&2618&1200&1200&yes\\
WMAP3 J1357+7644&---&S51357+769      &13 57 55.3&+76 43
19.9&Blazar Unknown-type&0&844&800&800&no\\ WMAP3
J1408$-$0749&WMAP 203&PKS1406-076     &14 08 56.4&$-$07 52
27.0&FSRQ&1.494&1080&1100&1100&yes\\ WMAP3
J1411+5216&---&3C295           &14 11 20.5&+52 12 09.0&Radio Galaxy&0.464&7374&0&0&no\\ WMAP3 J1419+3823&WMAP
42&B31417+385      &14 19 46.6&+38 21 47.9&FSRQ&1.831&651&1000&1000&yes\\ WMAP3
J1420+2702&---&GB6J1419+2706   &14 19 59.2&+27 06 24.9&FSRQ&0.536&415&1100&1000&yes\\ WMAP3 J1419+5425&---&1Jy1418+546
&14 19 46.6&+54 23 13.9&BL Lac&0.153&1350&900&900&no\\
WMAP3 J1427$-$3303&WMAP 193&PKS1424-328     &14 27 41.4&$-$33 05
30.9&Unidentified &0&352&1700&1600&yes\\ WMAP3 J1427$-$4206&WMAP
191&PKS1424-41      &14 27 56.2&$-$42 06 19.0&FSRQ&1.522&2597&2900&2900&yes\\ WMAP3
J1440+4958&---&GB6J1439+4958   &14 39 47.8&+49 58 05.0&BL Lac&0&198&0&0&no\\ WMAP3 J1442+5156&---&3C303           &14 43
02.7&+52 01 36.9&SSRQ&0.141&1070&800&800&no\\
WMAP3 J1458+7140&WMAP 71&3C309.1         &14 59 07.6&+71 40
18.9&SSRQ&0.905&3567&1100&1100&yes\\ WMAP3
J1504+1029&WMAP 6&PKS1502+106     &15 04 24.9&+10 29 38.0&FSRQ&1.839&2325&1600&1600&yes\\ WMAP3
J1506$-$1644&---&PKS1504-167     &15 07 04.8&$-$16 52 30.0&FSRQ&0.876&2840&1000&1000&yes\\ WMAP3 J1510$-$0548&---&4C-05.64
&15 10 53.5&$-$05 43 06.9&FSRQ&1.185&1742&1400&1400&yes\\
WMAP3 J1512$-$0905&WMAP 207&PKS1510-08      &15 12 50.5&$-$09 05
59.9&FSRQ&0.36&3080&1800&1800&yes\\ WMAP3
J1514$-$1013&---&PKS1511-10      &15 13 44.8&$-$10 11 59.9&FSRQ&1.513&1220&1100&1100&yes\\ WMAP3 J1516+0013&WMAP 2&4C00.56
&15 16 40.1&+00 15 02.0&Radio Galaxy&0.052&1641&1700&1700&yes\\ WMAP3 J1517$-$2421&WMAP
205&APLIB           &15 17 41.8&$-$24 22 18.9&BL Lac&0.049&2013&1900&1900&yes\\ WMAP3 J1540+1446&---&4C14.60
&15 40 49.5&+14 47 45.9&BL Lac&0.605&1210&800&800&no\\
WMAP3 J1549+0236&WMAP 5&PKS1546+027     &15 49 29.4&+02 37
00.9&FSRQ&0.414&1078&2100&2100&yes\\ WMAP3
J1549+5037&---&S41547+507      &15 49 17.5&+50 38 05.9&FSRQ&2.175&731&900&900&no\\ WMAP3 J1550+0526&WMAP 7&4C05.64
&15 50 35.2&+05 27 10.0&FSRQ&1.422&1766&2000&2000&yes\\
WMAP3 J1556$-$7912&---&PKS1549-79      &15 56 58.9&$-$79 14
03.9&Radio Galaxy&0.15&4676&500&500&no\\ WMAP3
J1601+3329&---&4C33.38         &16 02 07.2&+33 26 53.0&Blazar
Unknown-type/GPS&1.1&1656&900&900&no\\ WMAP3 J1608+1028&WMAP 9&4C10.45
&16 08 46.0&+10 29 07.0&FSRQ&1.226&1412&2200&2200&yes\\
WMAP3 J1617$-$7716&WMAP 183&PKS1610-77      &16 17 49.3&$-$77 17
17.9&FSRQ&1.71&3350&2000&1900&yes\\ WMAP3 J1613+3412&WMAP
23&OS319           &16 13 40.9&+34 12 47.0&FSRQ&1.397&2324&3300&3300&yes\\ WMAP3 J1635+3807&WMAP 33&4C38.41
&16 35 15.4&+38 08 03.9&FSRQ&1.813&3221&4900&4900&yes\\
WMAP3 J1637+4714&---&4C47.44         &16 37 45.1&+47 17
32.9&FSRQ&0.74&1244&900&900&no\\ WMAP3
J1643$-$7712&---&PKS1637-77      &16 44 16.0&$-$77 15 47.9&Radio Galaxy&0.043&2268&800&800&no\\ WMAP3 J1638+5722&WMAP
56&S41637+574      &16 38 13.3&+57 20 24.0&FSRQ&0.75&1750&1500&1500&yes\\ WMAP3 J1633+8227&WMAP 76&NGC6251
&16 32 26.1&+82 32 20.0&Radio Galaxy&0.024&1460&1400&1400&yes\\ WMAP3 J1642+3948&WMAP
35&3C345           &16 42 58.6&+39 48 37.0&FSRQ&0.593&8719&6000&6000&yes\\ WMAP3 J1642+6853&WMAP 69&4C69.21
&16 42 07.8&+68 56 38.0&FSRQ&0.751&1527&900&900&no\\
WMAP3 J1648+4114&---&GB6J1648+4104   &16 48 29.2&+41 04 05.0&SSRQ&0.851&197&1200&900&no\\ WMAP3 J1651+0457&WMAP
10&HerA            &16 51 08.1&+04 59 32.9&Radio Galaxy&0.154&11376&1100&1100&yes\\ WMAP3 J1654+3939&WMAP
36&MKR501          &16 53 52.2&+39 45 35.9&BL Lac&0.034&1375&0&0&no\\ WMAP3 J1658+0742&WMAP 13&PKS1655+077
&16 58 08.8&+07 41 27.9&FSRQ&0.621&1108&1000&1000&yes\\
WMAP3 J1656+5706&---&4C57.28         &16 57 20.7&+57 05
53.0&FSRQ&1.281&764&600&600&no\\ WMAP3
J1657+4749&---&S41656+47       &16 57 54.5&+47 49 14.9&FSRQ&0&0&600&600&no\\ WMAP3 J1703$-$6213&WMAP 198&PMNJ1703-6212
&17 03 36.4&$-$62 12 39.9&FSRQ&0&616&2000&2000&no\\ WMAP3
J1659+6829&---&GB6J1700+6830   &17 00 09.3&+68 30 06.0&FSRQ&0.301&380&700&700&no\\ WMAP3 J1707+0146&---&PKS1705+018
&17 07 34.3&+01 48 45.0&FSRQ&2.568&463&900&900&no\\ WMAP3
J1716+6840&---&S41716+68       &17 16 13.9&+68 36 38.0&FSRQ&0.777&838&700&700&no\\ WMAP3 J1724$-$6500&WMAP 196&NGC6328
&17 23 41.0&$-$65 00 36.0&Radio Galaxy/GPS&0.014&4373&1700&1700&yes\\ WMAP3 J1737$-$7934&WMAP
186&PKS1725-795     &17 33 40.6&$-$79 35 54.9&Blazar
Unknown-type&0&419&1100&1100&yes\\ WMAP3 J1727+4530&WMAP
43&S41726+45       &17 27 27.6&+45 30 38.9&FSRQ&0.717&935&1100&1100&yes\\ WMAP3 J1734+3856&WMAP
38&1Jy1732+389     &17 34 20.5&+38 57 51.0&FSRQ&0.97&557&1200&1200&yes\\ WMAP3 J1740+4739&---&S41738+476
&17 39 57.1&+47 37 58.0&BL Lac&0&818&1100&1100&yes\\ WMAP3
J1740+5212&WMAP 48&4C51.37         &17 40 36.9&+52 11 43.0&FSRQ&1.375&1699&1300&1300&yes\\ WMAP3 J1748+7006&WMAP
68&1Jy1749+701     &17 48 32.8&+70 05 50.9&BL Lac&0.77&715&900&900&no\\ WMAP3 J1753+2848&WMAP
22&GB6J1753+2847   &17 53 42.4&+28 48 03.9&Blazar
Unknown-type&0&511&2400&2400&yes\\ WMAP3 J1753+4405&---&S41751+441
&17 53 22.6&+44 09 46.0&FSRQ&0.871&1000&800&800&no\\
WMAP3 J1803$-$6507&WMAP 199&PKS1758-651     &18 03 23.5&$-$65 07
36.0&Blazar Unknown-type&0&780&1500&1500&yes\\ WMAP3
J1758+6632&WMAP 64&NGC6543         &17 58 33.4&+66 37
58.0&Planetary Nebula&0&875&600&600&no\\ WMAP3
J1759+3852&---&S41758+3        &18 00 24.7&+38 48 29.9&FSRQ&2.092&735&800&800&no\\ WMAP3 J1801+4404&---&1Jy1800+440
&18 01 32.2&+44 04 22.0&FSRQ&0.663&1193&1400&1400&yes\\
WMAP3 J1800+7828&WMAP 72&S51803+784      &18 00 45.7&+78 28
04.0&BL Lac&0.68&2620&1700&1700&yes\\ WMAP3 J1806+6949&WMAP
67&3C371           &18 06 50.3&+69 49 27.9&BL Lac&0.051&2122&1200&1200&yes\\ WMAP3 J1820$-$6343&WMAP
200&PKS1814-63      &18 19 34.9&$-$63 45 47.9&Radio Galaxy&0.063&4506&1200&1200&yes\\ WMAP3
J1820$-$5521&---&PKS1815-553     &18 19 45.4&$-$55 21 20.0&FSRQ&0&1124&0&0&no\\ WMAP3 J1824+5650&WMAP 53&4C56.27
&18 24 07.0&+56 51 01.0&BL Lac&0.664&1263&1400&1400&yes\\
WMAP3 J1829+4844&WMAP 46&3C380.0         &18 29 31.7&+48 44
45.9&SSRQ&0.692&5519&2500&2500&yes\\ WMAP3
J1837$-$7106&WMAP 192&PKS1831-711     &18 37 28.6&$-$71 08
43.0&FSRQ/GPS&1.356&2294&1500&1500&yes\\ WMAP3
J1835+3246&---&3C382.0         &18 35 02.1&+32 41 49.9&Radio Galaxy&0.058&2281&700&700&no\\ WMAP3 J1842+6807&WMAP
66&S41842+68       &18 42 33.6&+68 09 24.0&FSRQ&0.472&925&1200&1200&yes\\ WMAP3 J1840+7946&WMAP 73&3C390.3
&18 42 09.0&+79 46 17.0&SSRQ&0.056&4380&1000&1000&yes\\ WMAP3
J1848+3221&---&B21846+32A      &18 48 22.0&+32 19 01.9&FSRQ&0.798&762&800&800&no\\ WMAP3 J1850+2823&WMAP 28&S21848+28
&18 50 27.4&+28 25 12.0&FSRQ/GPS&2.56&1013&900&900&no\\ WMAP3
J1849+6705&WMAP 65&4C66.20         &18 49 15.9&+67 05 39.9&FSRQ&0.657&845&1400&1400&yes\\ WMAP3 J1902+3153&WMAP 34&3C395
&19 02 55.9&+31 59 42.0&FSRQ&0.635&1863&800&800&no\\
WMAP3 J1923$-$2106&WMAP 8&PMNJ1923-2104   &19 23 32.2&$-$21 04
32.9&FSRQ&0.874&2885&2100&2100&yes\\ WMAP3
J0000+0000&---&OV-236          &19 24 51.4&$-$29 14 30.0&FSRQ&0.352&14332&13200&13200&yes\\ WMAP3 J1927+6118&WMAP
59&S41926+611      &19 27 30.4&+61 17 31.9&BL Lac&0&558&1100&1100&yes\\ WMAP3 J1927+7357&WMAP 70&4C73.18
&19 27 48.4&+73 58 00.9&FSRQ&0.302&3626&2900&2900&yes\\
WMAP3 J1937$-$3956&---&PKS1933-400     &19 37 16.3&$-$39 58
00.9&FSRQ&0.965&1129&1100&1100&yes\\ WMAP3
J1938$-$6345&---&PKS1934-63      &19 39 25.0&$-$63 42 44.9&Blazar
Unknown-type/GPS&0.183&5910&0&0&no\\ WMAP3 J1952+0233&---&3C403.0
&19 52 15.7&+02 30 24.0&Radio Galaxy&0.059&2026&600&600&no\\ WMAP3 J1955+5132&WMAP
51&S41954+513      &19 55 42.7&+51 31 49.0&FSRQ&1.22&1675&800&800&no\\ WMAP3 J1958$-$3845&WMAP
3&PKS1954-388     &19 57 59.8&$-$38 45 06.0&FSRQ&0.63&1944&3000&3000&yes\\ WMAP3 J2000$-$1749&WMAP
11&PKS1958-179     &20 00 57.1&$-$17 48 56.9&FSRQ&0.65&2700&1600&1600&yes\\ WMAP3 J2005+7755&---&S52007+777
&20 05 31.1&+77 52 42.9&BL Lac&0.342&1260&0&0&no\\ WMAP3
J2011$-$1547&WMAP 14&PKS2008-159     &20 11 15.7&$-$15 46
40.0&FSRQ&1.18&949&1500&1500&yes\\ WMAP3
J2010+7230&---&4C72.28         &20 09 52.4&+72 29 18.9&BL Lac&0&907&800&800&no\\ WMAP3 J2022+6136&WMAP 63&S42021+61
&20 22 06.7&+61 36 57.9&Radio Galaxy&0.227&2743&1300&1300&no\\ WMAP3 J2035$-$6845&WMAP
194&PKS2030-689     &20 35 48.7&$-$68 46 32.9&FSRQ&1.084&648&800&800&no\\ WMAP3 J2035+1058&---&PKS2032+107
&20 35 22.3&+10 56 07.0&BL Lac&0.601&759&0&0&no\\ WMAP3
J2056$-$4717&WMAP 208&PKS2052-47      &20 56 16.3&$-$47 14
47.0&FSRQ&1.489&2026&1900&1900&yes\\ WMAP3
J2109$-$4114&WMAP 1&PKS2106-413     &21 09 33.1&$-$41 10
19.9&FSRQ&1.058&2076&1200&1200&yes\\ WMAP3
J2109+3537&WMAP 49&GB6J2109+3532   &21 09 31.9&+35 32
57.0&Unidentified &0&1208&700&700&no\\ WMAP3 J2123+0536&WMAP
27&PKS2121+053     &21 23 44.5&+05 35 21.9&FSRQ&1.878&1930&1700&1700&yes\\ WMAP3 J2131$-$1206&WMAP
17&PKS2128-12      &21 31 35.2&$-$12 07 05.9&FSRQ&0.501&2989&2300&2300&yes\\ WMAP3 J2134$-$0154&WMAP
20&4C-02.81        &21 34 10.3&$-$01 53 17.0&BL Lac&1.285&1308&1700&1700&yes\\ WMAP3 J2136+0042&WMAP
25&PKS2134+004     &21 36 38.5&+00 41 53.9&FSRQ&1.932&8416&3000&3000&yes\\ WMAP3 J2139+1424&WMAP
41&PKS2136+141     &21 39 01.3&+14 23 35.0&FSRQ&2.427&1073&1900&1900&yes\\ WMAP3 J2148$-$7757&WMAP
184&PKS2141-781     &21 46 29.9&$-$77 55 54.0&FSRQ&0.334&508&0&0&no\\ WMAP3 J2143+1740&WMAP 44&S32141+17
&21 43 35.5&+17 43 49.0&FSRQ&0.211&1006&0&0&no\\ WMAP3
J2148+0657&WMAP 37&4C06.69         &21 48 05.3&+06 57 38.9&FSRQ&0.99&4949&7200&7200&yes\\ WMAP3
J2151$-$3026&---&PKS2149-307     &21 51 55.4&$-$30 27 52.9&FSRQ&2.345&1336&1200&1200&yes\\ WMAP3 J2157$-$6942&WMAP
190&PKS2153-69      &21 57 05.9&$-$69 41 23.9&SSRQ&0.028&11649&2600&2600&yes\\ WMAP3 J2158$-$1502&WMAP
18&PKS2155-152     &21 58 06.1&$-$15 01 09.0&FSRQ&0.672&3091&1700&1700&yes\\ WMAP3 J2202+4217&WMAP 58&BLLAC
&22 02 43.2&+42 16 40.0&BL Lac&0.069&2940&3100&3100&no\\
WMAP3 J2203+3146&WMAP 54&4C31.63         &22 03 15.0&+31 45
38.0&FSRQ&0.295&2806&2200&2200&yes\\ WMAP3
J2203+1724&WMAP 45&PKS2201+171     &22 03 26.8&+17 25 47.9&FSRQ&1.075&834&1700&1700&yes\\ WMAP3 J2206$-$1839&WMAP
16&PKS2203-18      &22 06 10.3&$-$18 35 39.0&FSRQ&0.619&4271&1300&1300&yes\\ WMAP3
J2207$-$5349&---&PKS2204-54      &22 07 43.6&$-$53 46 33.9&FSRQ&1.206&1410&800&800&no\\ WMAP3 J2211+2351&WMAP
50&PKS2209+236     &22 12 05.9&+23 55 40.0&FSRQ/GPS&1.125&1123&1400&1400&yes\\ WMAP3 J2218$-$0335&WMAP
30&4C-03.79        &22 18 52.0&$-$03 35 36.9&FSRQ&0.901&2717&2000&2000&yes\\ WMAP3 J2225$-$0456&WMAP 29&3C446
&22 25 47.2&$-$04 57 01.0&FSRQ&1.404&6382&4100&4100&yes\\
WMAP3 J2229$-$0832&WMAP 24&PKS2227-08      &22 29 40.0&$-$08 32
53.9&FSRQ&1.559&2423&2600&2600&yes\\ WMAP3
J2229$-$2051&---&PMNJ2229-2049   &22 29 48.1&$-$20 49
26.0&Planetary Nebula&0&326&800&800&no\\ WMAP3 J2232+1144&WMAP
47&4C-11.69        &22 32 36.4&+11 43 50.0&FSRQ&1.037&3967&3400&3400&yes\\ WMAP3 J2235$-$4834&WMAP
206&PKS2232-488     &22 35 13.1&$-$48 35 57.9&FSRQ&0.51&1104&1800&1800&yes\\ WMAP3 J2236+2825&WMAP
57&B22234+28A      &22 36 22.3&+28 28 58.0&FSRQ&0.795&1625&1100&1100&yes\\ WMAP3 J2239$-$5700&WMAP
201&PKS2236-572     &22 39 12.1&$-$57 01 00.0&Unidentified
&0&1063&0&0&no\\ WMAP3 J2246$-$1210&WMAP 21&PKS2243-123     &22 46
18.1&$-$12 06 51.9&FSRQ&0.632&2661&1100&1100&yes\\ WMAP3
J2254+1608&WMAP 55&3C454.3         &22 53 57.7&+16 08 53.0&FSRQ&0.859&14468&6800&6800&yes\\ WMAP3 J2255+4202&---&S42253+41
&22 55 36.7&+42 02 50.9&FSRQ&1.476&1084&900&800&no\\
WMAP3 J2256$-$2011&WMAP 19&PKS2254-204     &22 56 41.0&$-$20 11
40.9&BL Lac&0&454&800&800&no\\ WMAP3 J2258$-$2757&WMAP
12&PKS2255-282     &22 58 05.8&$-$27 58 18.9&FSRQ&0.926&2127&6600&6600&yes\\ WMAP3 J2315$-$5017&WMAP
204&PKS2312-505     &23 15 44.2&$-$50 18 38.9&Unidentified
&0&341&900&900&no\\ WMAP3 J2322+5107&---&GB6J2322+5057   &23 22
25.9&+50 57 51.0&FSRQ&1.279&1305&800&800&no\\ WMAP3
J2327+0939&---&PKS2325+093     &23 27 33.4&+09 40 09.0&FSRQ&1.843&886&1200&1200&yes\\ WMAP3
J2329$-$4732&---&PKS2326-477     &23 29 17.7&$-$47 30 19.0&FSRQ&1.299&2206&1400&1400&yes\\ WMAP3 J2330+1056&---&4C10.73
&23 30 40.7&+11 00 18.0&FSRQ&1.489&1194&1000&1000&yes\\
WMAP3 J2331$-$1559&WMAP 32&PKS2329-16      &23 31 38.5&$-$15 56
57.0&FSRQ&1.153&1879&800&800&no\\ WMAP3
J2333$-$2340&---&PKS2331-240     &23 33 55.2&$-$23 43 40.0&Radio Galaxy&0.048&907&1200&1200&yes\\ WMAP3
J2334+0734&---&GB6J2334+0736   &23 34 12.7&+07 36 25.9&FSRQ&0.401&953&1100&1100&yes\\ WMAP3
J2335$-$0129&---&PKS2332-017     &23 35 20.3&$-$01 31 09.0&FSRQ&1.184&581&1300&1300&yes\\ WMAP3 J2335$-$5242&WMAP
195&PKS2333-528     &23 36 12.1&$-$52 36 21.9&FSRQ&0&1588&500&400&no\\ WMAP3 J2346+0927&---&4C09.74
&23 46 36.7&+09 30 44.9&FSRQ&0.677&1344&700&600&no\\
WMAP3 J2348$-$1631&WMAP 39&PKS2345-16      &23 48 02.5&$-$16 31
11.9&FSRQ&0.6&2360&2000&2000&yes\\ WMAP3 J2354+4550&WMAP
74&4C45.51         &23 54 21.7&+45 53 03.9&FSRQ&1.992&1127&1300&1300&yes\\ WMAP3 J2356+4952&WMAP
75&1Jy2352+495     &23 55 09.4&+49 50 08.0&Radio Galaxy&0.238&2306&0&0&no\\ WMAP3 J2357$-$5313&WMAP
189&PKS2355-534     &23 57 53.2&$-$53 11 12.9&FSRQ&1.006&1782&1100&1100&yes\\ WMAP3
J2358$-$1013&---&PKS2355-106     &23 58 10.8&$-$10 20 08.0&FSRQ&1.639&1618&0&0&no\\ WMAP3 J2358$-$6050&WMAP 187&PKS2356-61
&23 58 45.5&$-$60 52 54.9&Radio Galaxy&0.096&4421&1200&1200&yes\\

 \label{tab:list}
\end{longtable}
\end{landscape}
\end{document}